\documentclass[12pt, a4paper]{article}

\usepackage[french,english]{babel}
\usepackage[T1]{fontenc}

\usepackage{mathptmx} 
\usepackage{helvet}   

\usepackage{amsmath}
\usepackage{amssymb}

\usepackage{graphicx}
\usepackage{booktabs}
\usepackage{multirow}
\usepackage{array}

\usepackage{tikz}
\usetikzlibrary{arrows,shapes,positioning,shadows,trees}
\usepackage{pgfplots}

\usepackage{cite}
\usepackage{url}
\usepackage{graphicx}
\usepackage{footnote}
\makesavenoteenv{tabular}
\makesavenoteenv{table}

\title{Résumé abstractif à partir d'une transcription audio} 
\author{Ilia Derkach\\Université Grenoble Alpes\\Laboratoire d'Informatique de Grenoble\\150 place du Torrent, 38401 Saint Martin d'Hères}
\date{} 

\begin{document}
\maketitle
\begin{abstract} 
Les modèles de langue de grande taille gagnent actuellement en popularité et leurs résultats sont utilisés dans de nombreux domaines, de la traduction de texte à la génération de réponses aux requêtes. Cependant, le principal problème de ces nouveaux algorithmes d'apprentissage automatique réside dans le fait que leur apprentissage nécessite d'importantes ressources de calcul. Pour pallier ce problème, plusieurs méthodes (LoRA, quantification) ont été proposées afin d'affiner efficacement les modèles existants pour des tâches spécifiques. Dans cet article, nous proposons un modèle de résumé audio de bout en bout (E2E) utilisant ces techniques. De plus, cet article examine l'efficacité de ces approches pour le problème étudié et tire des conclusions sur leur applicabilité.
\end{abstract}

\section{Introduction}

Dans le monde moderne, divers modèles d'apprentissage automatique sont largement utilisés dans de nombreux aspects de notre vie quotidienne. Il est devenu courant d'utiliser de tels modèles non seulement pour prédire des classes ou des valeurs numériques (comme dans un problème de régression), mais aussi pour prédire et générer des objets plus complexes tels que du texte, de l'audio ou même de la vidéo. La tâche de résumé audio se distingue parmi celles-ci. Elle consiste à créer un modèle capable de générer une brève description de ce qui a été dit à partir d'un enregistrement vocal. Alors qu'autrefois, même la tâche de résumer du texte semblait difficile et que les modèles ne produisaient pas de résultats comparables à une évaluation humaine, l'avènement des grands modèles de langage a rendu cette tâche réalisable.

Il est important de noter que cette tâche n'est pas seulement intéressante sur le plan théorique. Un tel modèle peut être intégré dans de nombreuses applications modernes utilisées quotidiennement par des millions de personnes. Par exemple, ce type de modèle pourrait être utilisé pour générer automatiquement des descriptions vidéo sur des plateformes populaires, ou les résumés générés pourraient servir de notifications push pour les messages audio dans des applications de messagerie courantes. Une telle solution pourrait également faire gagner beaucoup de temps aux employés d'entreprises qui passent une grande partie de leur journée en réunions en ligne. À la fin de chaque session, les résultats, décisions et plans formulés pourraient être générés par le modèle de résumé audio.

Bien qu'il ait été mentionné que l'apparition des grands modèles de langage a permis une avancée significative dans l'étude de nombreuses tâches, notamment celle du résumé audio, ces modèles présentent également des inconvénients. En raison de leur grand nombre de paramètres et de leur architecture complexe, l'entraînement complet de tels modèles est chronophage et nécessite des ressources informatiques importantes. Par exemple, Chat-GPT, le modèle de langage le plus connu, compte 175 milliards de paramètres et 570 Go de données d'entraînement. Bien que la durée d'entraînement de ce modèle soit officiellement inconnue, à titre de comparaison, le modèle Kandinsky (12 milliards de paramètres) a nécessité 20 352 jours-GPU-V100 pour son entraînement. Malheureusement, de telles limitations signifient qu'il est actuellement impossible de développer de tels modèles "à partir de zéro" avec les capacités d'un laboratoire de recherche. Cependant, il s'avère que les modèles déjà entraînés par de grandes entreprises ont un potentiel suffisant pour résoudre des tâches plus spécialisées. Grâce à leur grand nombre de paramètres et au volume important des ensembles de données d'entraînement, ces modèles assimilent bien la structure du langage, ce qui signifie que leur réajustement nécessite une puissance de calcul beaucoup plus faible lors de l'utilisation de diverses tactiques de réentraînement.

Le résumé de la parole en texte (S2T) utilise généralement une approche en cascade, où un modèle de reconnaissance automatique de la parole (ASR) génère des transcriptions, suivies par un modèle de résumé texte-texte (T2T) qui produit des résumés. Les avancées en apprentissage profond, en particulier les architectures basées sur l'attention et le pré-entraînement auto-supervisé, ont considérablement amélioré les performances des deux modèles. Les systèmes de résumé abstrait en cascade utilisant ces composants avancés fonctionnent bien sur les tâches de résumé de dialogue lorsqu'ils sont entraînés sur des données non appariées. Cependant, les transcriptions produites par le modèle ASR peuvent contenir des erreurs, ce qui a conduit au développement et à l'utilisation de méthodes telles que les réseaux de neurones profonds ou les modèles de langage pour améliorer la robustesse et atténuer l'impact de ces erreurs.

Une limitation majeure des systèmes en cascade pour le résumé S2T est leur incapacité à utiliser des informations non verbales et acoustiques, telles que l'intonation, les pauses et les émotions des locuteurs, qui pourraient fournir un contexte précieux pour un résumé plus précis. Pour remédier à ce problème, la modélisation de bout en bout (E2E) a été proposée. Les systèmes E2E contournent l'étape intermédiaire de reconnaissance vocale et optimisent conjointement les modèles acoustiques et linguistiques dans un cadre unifié. Cette approche peut potentiellement capturer des informations plus riches directement à partir du signal audio, conduisant à des résumés plus précis et contextuellement pertinents.

Cependant, la modélisation E2E présente ses propres défis. Elle nécessite de grandes quantités de données audio/résumé appariées pour un entraînement efficace, ce qui est souvent rare, en particulier dans des domaines spécifiques comme les actualités télévisées. La disponibilité limitée de grands corpus publics nécessite le développement de techniques pour exploiter des sources de données externes. Par exemple, des stratégies d'apprentissage par transfert et d'augmentation des données peuvent être employées pour améliorer les performances du modèle. De plus, les approches d'apprentissage non supervisé et semi-supervisé peuvent aider à mieux utiliser les ensembles de données non appariées ou partiellement appariées, répondant ainsi au problème de la rareté des données.

Ce travail propose un modèle E2E pour le résumé abstrait S2T. Le modèle utilise un ASR et un modèle de résumé abstrait T2T affinés sur un ensemble de données de vidéos descriptives, en utilisant des méthodes telles que LoRA et la quantification. Le système E2E suit le paradigme encodeur-décodeur et utilise des caractéristiques vocales extraites des formes d'onde fournies dans l'ensemble de données. Comme mentionné précédemment, en raison des spécificités du modèle, un modèle de résumé de texte et un modèle de reconnaissance vocale ont été autorisés. Les métriques calculées pour ces modèles ont été comparées avec d'autres résultats classiques dans ces domaines, ce qui nous a permis de tirer des conclusions sur l'applicabilité des méthodes d'affinage décrites ci-dessus.

Ce document est organisé comme suit : dans la section 2, je présente la description des méthodes d'affinage efficaces des grands modèles de langage telles que LoRA et la quantification, qui ont été utilisées dans les expériences ; dans la section 3, je décris l'ensemble de données sur lequel l'entraînement a été effectué et l'architecture du modèle de résumé abstrait T2T, ainsi que la présentation des résultats et la comparaison avec les modèles existants ; dans la section 4, la même chose est décrite pour le modèle ASR ; dans la section 5, je présente comment les modèles entraînés peuvent être combinés en un seul modèle E2E de parole en texte, et je décris également les plans pour les recherches futures.\section{Etat de l'art sur ajustement fin et efficace des paramètres}
Dans le domaine en constante évolution de l'apprentissage automatique, la quête de modèles plus efficaces et évolutifs a conduit à des innovations significatives dans les techniques d'affinage des modèles. Le réglage fin et efficace des paramètres (\textit{Parameter Efficient Fine-Tuning} - PEFT) s'est imposé comme une approche cruciale, permettant d'adapter de grands modèles pré-entraînés à des tâches spécifiques avec une fraction du coût et des ressources de calcul. Ce chapitre explore la bibliothèque PEFT \cite{19}, en se concentrant plus particulièrement sur l'adaptation de bas rang (\textit{Low Rank Approximation} - LoRA) \cite{13, 14, 15} et les méthodes de quantification \cite{17, 18}, que j'ai utilisées pour affiner les modèles destinés à mes tâches spécifiques.

\subsection{Low-Rank Adaptation (LoRA)}

\begin{figure}[!ht]
\centering
\includegraphics[width=0.4\textwidth]{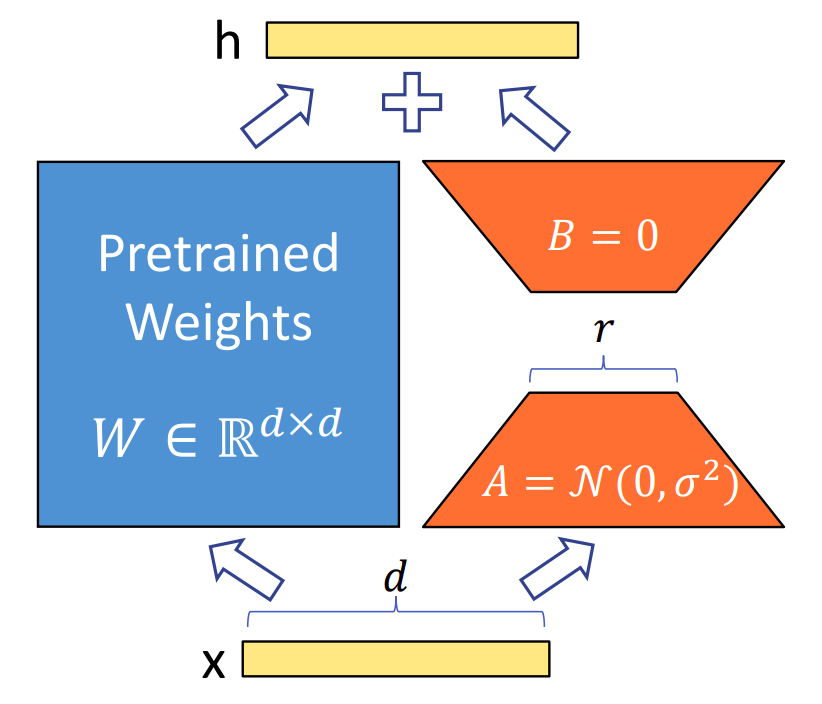}
\caption{Représentation schématique de LoRA. Les pondérations $W$ sont figées, tandis que $\Delta W = A\cdot B$}
\label{fig:lora}
\end{figure}

Pour illustrer le fonctionnement de ce mécanisme, considérons une couche linéaire ordinaire représentée comme suit :
\[
y = Wx,
\]
où $x$ est l'entrée de la couche et $W$ la matrice des pondérations. Lors de l'entraînement ultérieur du modèle, il est nécessaire de modifier légèrement le fonctionnement de cette couche en ajustant les pondérations de $\Delta W$ (généralement recherchées par la descente de gradient), de sorte que la nouvelle sortie soit la suivante :
\[
y' = W'x = (W + \Delta W)x = Wx + \Delta Wx
\]
Comme on peut le constater, la nouvelle valeur de $y$ diffère de l'ancienne de $\Delta W$, ce qui peut être interprété comme le résultat du travail d'une seule couche distincte et entièrement connectée. Cette interprétation est illustrée dans la figure \ref{fig:lora} ci-dessous.
Ainsi, les pondérations de la matrice $W$ peuvent être fixes, et celles de $\Delta W$ peuvent être entraînées. Cela peut être interprété comme l'entraînement d'un modèle prédisant la différence entre le résultat d'un modèle conventionnel et celui d'un modèle pré-entraîné. Cela rappelle vaguement le gradient boosting, où chaque arbre de décision suivant est entraîné pour corriger les erreurs du modèle précédent.

Cependant, une question légitime peut se poser : pourquoi cette transformation est-elle une optimisation si le nombre de pondérations dans $W$ et $\Delta W$ est le même ? Afin de réduire le nombre de pondérations, la décomposition matricielle en produit de deux matrices de rang inférieur est utilisée, comme indiqué dans le nom de cette méthode. Plus précisément, la matrice $\Delta W$ de taille $n\times k$ est représentée par $A\cdot B$, où $A$ et $B$ ont des tailles $n\times r$ et $r\times k$, où $r$ est une petite constante. Dans l'article original \cite{13}, il a été démontré qu'il suffit de choisir un $r$ relativement petit avec un rang compris entre 8 et 128, auquel cas le nombre de paramètres entraînés sera inférieur à
\[
\frac{n \cdot k}{(n + k) \cdot r} \approx 10^2, \text{en général.}
\]
La question peut se poser de savoir si le modèle perdra en complexité généralisante si nous choisissons un paramètre $r$ suffisamment petit. Cependant, les auteurs de l'article \cite{13} ont démontré que dans la réalité, le « rang interne » des grands modèles de texte est très faible et que la plupart des paramètres n'ont pas beaucoup d'impact.

Lors de l'initialisation du modèle, la matrice B est définie aléatoirement (par exemple, à partir de $\mathcal{N}(0, \sigma^2)$), et la matrice $A$ est initialisée avec des zéros, de sorte qu'initialement $\Delta W = 0$.

Ces couches peuvent être perçues comme des adaptateurs de couches linéaires. Si l'on considère l'architecture du transformateur (qui, comme nous le verrons plus loin, inclut les modèles E2E et ASR en cours d'apprentissage), ces couches sont généralement appliquées au poids de la matrice d'auto-attention. Plus précisément, le module d'auto-attention est représenté par :
\[
Auto-attention(Q, K, V) = softmax\left(\frac{QK^T}{\sqrt{d_k}} \right)V,
\]
où $d_k$ est la dimension des clés et des requêtes (une des dimensions des matrices $Q$ et $K$) et les matrices $Q$ (requêtes), $K$ (clés) et $V$ (valeurs) sont des projections linéaires. de l'entrée $X$ :
\[
Q = XW^Q \;\;\;\;\;\;\;\;\; K = XW^K \;\;\;\;\;\;\;\;\;\; V = XW^V,
\]
où $W^Q$, $W^K$, $W^V$ sont des matrices de projection apprises. La méthode LoRA sera ensuite appliquée aux matrices $W^Q$, $W^K$, $W^V$.

Bien que cette méthode soit relativement efficace en termes d'utilisation mémoire et de temps d'apprentissage, elle présente un inconvénient majeur. Les transformateurs possèdent de nombreuses couches d'attention auxquelles LoRA est appliqué. Cependant, le rang $r$ de la décomposition est le même pour tous les adaptateurs, ce qui est certainement inefficace : il est clair que certaines matrices contribuent davantage et ont un rang élevé, tandis que d'autres ont un rang mineur. Cet inconvénient a été corrigé dans une version améliorée de l'algorithme appelée AdaLoRA \cite{14}, qui sera abordée dans la section suivante.

\subsection{Allocation budgétaire adaptative pour un réglage fin efficace des paramètres (AdaLoRA)}

Pour corriger l'inefficacité décrite dans la section précédente, la méthode AdaLoRA \cite{14} a été proposée. Elle répartit de manière adaptative le budget de paramètres entre les matrices de pondération en fonction de leur score d'importance. AdaLoRA paramètre les adaptateurs à l'aide d'une forme de décomposition en valeurs singulières (SVD) \cite{20}. Cette approche innovante permet d'éliminer efficacement les valeurs singulières des mises à jour moins importantes, réduisant ainsi leur budget de paramètres tout en évitant les calculs SVD exacts et intensifs. Pour entrer dans les détails, supposons que l'adaptateur LoRA $\Delta W$ présente la décomposition en valeurs singulières suivante :
\[
\Delta W = P\Lambda Q,
\]
où $P \in \mathbb{R}^{d_1 \times r}$ et $Q \in \mathbb{R}^{r \times d_2}$ représentent les vecteurs singuliers gauche/droite de $\Delta W$ et $\Lambda \in \mathbb{R}^{r \times r}$ est une matrice diagonale à valeurs singulières, correspondant respectivement aux vecteurs singuliers gauche/droite des matrices $P$ et $Q$. Alors que la méthode précédente entraîne les matrices $A$ et $B$ de dimension arbitraire $r$, l'approche actuelle entraîne les matrices $P$, $Q$ et $\Lambda$ issues de SVD.

Il est important de comprendre qu'outre les modifications des matrices entraînées, le processus d'entraînement est également légèrement modifié. Si auparavant l'erreur d'apprentissage correspondait à un coût d'apprentissage $\mathcal{C}(P, \Lambda, Q)$, qui illustre l'écart entre la sortie du modèle actuel et la valeur cible, alors, dans l'algorithme actuel, un terme de régularisation est ajouté au coût d'apprentissage, conçu pour maintenir l'orthogonalité des matrices $P$ et $Q$. Formellement, la fonction de perte mise à jour pour chaque couche à laquelle AdaLoRAA est appliqué se présente comme suit :
\[
\mathcal{L}(P, \Lambda, Q) = \mathcal{C}(P, \Lambda, Q) + \gamma \begin{Vmatrix}P^TP - I\end{Vmatrix}_F^2 + \gamma \begin{Vmatrix}QQ^T - I\end{Vmatrix}_F^2
\]
De plus, le processus de mise à jour des pondérations entraînées est également modifié. Si rien ne change pour les matrices $P$ et $Q$, et que celles-ci continuent d'être mises à jour par descente de gradient stochastique sur la fonction de perte mise à jour, alors la mise à jour de $\Lambda$ change afin de modifier le rang de la matrice produit $\Delta W$. Plus précisément, pour la matrice $\Lambda$, le gradient stochastique est calculé comme suit (uniquement pour les éléments diagonaux, les autres éléments restant égaux à 0) :
\[
\tilde{\Lambda}^t = \Lambda^t - \eta \nabla \mathcal{L}(P^t, \Lambda^t, Q^t),
\]
où $\eta > 0$ est le taux d'apprentissage par descente de gradient. Cependant, les valeurs singulières de $\tilde{\Lambda}^t$ sont ensuite élaguées en fonction du score d'importance $S^t$ :
\[
\Lambda^{t + 1}_{i, i} = \begin{cases}
\tilde{\Lambda}^t_{i, i} & \text{si $S^t_i$ est dans le top-$b^t$ de $S_t$} \\
0 & \text{sinon},
\end{cases}
\]
où $S^t$ contient le score d'importance de toutes les valeurs singulières et de ses vecteurs singuliers gauche/droite, et $b^t$ est le budget des valeurs singulières restantes à la $t$-ième étape. $b^t$ doit être considéré comme une fonction décroissante dépendant de $t$, ce qui permet d'éliminer efficacement la proportion de nombres singuliers sans importance. Dans le même temps, dans l'exemple le plus simple, $S^t$ peut ne dépendre que de valeurs singulières et être égal à sa valeur absolue : $S^t_i = |\tilde{\lambda}^t_i|$ (une valeur singulière élevée est plus importante à préserver lors de l'élagage). En réalité, il a été démontré qu'il est judicieux de choisir une fonction plus complexe dépendant à la fois des valeurs singulières et des vecteurs singuliers correspondants.
\subsection{Quantification}

La quantification est le processus de conversion de valeurs d'une représentation riche en informations (généralement un ensemble continu) vers une représentation plus compacte, généralement un ensemble discret. L'échantillonnage de signaux est un bon exemple de quantification : à chaque valeur d'un signal continu est attribuée une valeur issue d'un ensemble discret prédéfini. Dans le contexte des réseaux de neurones, la quantification consiste à passer d'un type de données comportant un grand nombre de bits, tel que \texttt{float32}, à un type comportant un nombre plus petit, tel que \texttt{int8}. Cette approche permet de réduire considérablement le poids du modèle sans altérer sa qualité. Cela peut s'avérer crucial, par exemple, pour les modèles exécutés directement sur des appareils mobiles. Concernant les modèles LLM de pointe, il est impossible de les enregistrer directement sur le téléphone, dans l'application mobile, car la quantité de mémoire du modèle ne le permet pas. Si l'application elle-même ne permet pas aux développeurs de travailler avec le modèle sur le serveur, sans quantification, il sera impossible de l'enregistrer sur le téléphone. De plus, les modèles quantifiés nécessitent moins de ressources de calcul et s'exécutent plus rapidement. Le fonctionnement de cette transformation pour les pondérations des réseaux de neurones sera abordé plus en détail ci-après.

Il existe différentes approches de quantification. Cependant, seules les deux plus courantes seront considérées, car elles sont utilisées dans ce travail. La première, la plus simple et la plus intuitive, est la transformation linéaire. En termes généraux, cela signifie que la plage de valeurs initiale $[R_{\min}; R_{\max}]$ se transforme en plage quantifiée $[Q_{\min}; Q_{\max}]$ par transformation affine. On peut la diviser en deux types : la quantification asymétrique et la quantification symétrique. Considérons d'abord la transformation asymétrique.

Notons $S$ et $Z$ comme des constantes de quantification, c'est-à-dire des paramètres calculés au cours du processus. Définissez $S$ comme suit :
\[
S = \frac{R_{\max} - R_{\min}}{Q_{\max} - Q_{\min}}
\]
La constante $S$, appelée échelle, est stockée dans le type initial. La constante $Z$, appelée point zéro, définit le point où la valeur zéro initiale sera transformée. L'expression mathématique de $Z$ est :
\[
Z = \left[Q_{\min} - \frac{R_{\min}}{S} \right]
\]
Il est très important pour les réseaux de neurones de représenter précisément le zéro ; cette constante est donc déterminée. Dans la définition ci-dessus, l'arrondi peut être effectué de différentes manières : à l'entier inférieur ou supérieur, voire à l'entier le plus proche. $Z$, opposé à $S$, est souvent stocké dans un type quantifié. Les fonctions de quantification et de déquantification sont les suivantes :
\[
X_q = \left[\frac{X}{S} + Z \right] \;\;\;\;\;\; \text{et} \;\;\;\;\;\;\; X = S(X_q - Z).
\]
La quantification asymétrique est bien adaptée aux distributions non symétriques, par exemple pour la sortie ReLU.

La quantification symétrique fonctionne de manière similaire, à la différence que le point zéro est zéro, ce qui garantit la symétrie des valeurs par rapport à zéro. De plus, les limites de la plage quantifiée sont définies comme le maximum modulo la valeur quantifiée $|R_{\max}|$. Enfin, pour que le type soit symétrique, il faut ignorer une valeur du type de données quantifié. Par exemple, la plage \texttt{unsigned int8}: [-128, 127] deviendra [-127, 127]. Les constantes définies ci-dessus sont définies comme suit :
\[
S = \frac{|R_{\max}|}{2^{N - 1} - 1} \;\;\;\;\;\;\; \text{et} \;\;\;\;\;\;\; Z = 0,
\]
où $N$ est le nombre de bits de type quantifié. La quantification et la déquantification sont
\[
X_q = \left[\frac{X}{S} \right] \;\;\;\;\;\;\; \text{et} \;\;\;\;\;\;\; X = SX_q
\]
La deuxième approche de transformation des pondérations lors de la quantification repose sur le fait que les pondérations ne sont pas distribuées linéairement sur toute la plage, mais selon une distribution normale. On considérera ensuite que toutes les pondérations sont comprises dans la plage $[-1, 1]$. À cet égard, il est judicieux d'attribuer davantage de valeurs de type quantifié aux pondérations du milieu de la plage, et moins de valeurs aux pondérations des arêtes. Un exemple de cette approche est NF4, la construction d'un type de données 4 bits utilisé dans les expériences de cet article.

Il y a 4 bits, ce qui signifie que seules 16 valeurs peuvent être stockées. Deux valeurs doivent représenter -1 et 1 ; il n'en reste donc que 14. Elles servent à représenter des quantiles $\mathcal{N}(0, \sigma^2)$ dans l'intervalle [-1, 1]. Ensuite, chaque pondération peut être corrélée à la valeur la plus proche parmi les quantiles calculés précédemment. Cependant, le schéma décrit présente un inconvénient : il ne propose pas de représentation exacte de zéro. Comme mentionné précédemment, la valeur exacte de zéro est essentielle pour les réseaux de neurones, par exemple pour pouvoir effectuer des remplissages et d'autres éléments à valeur nulle sans erreur. Les auteurs proposent donc une solution élégante. Tout d'abord, l'intervalle [-1, 1] est divisé en deux parties, positive et négative. On trouve ensuite 7 quantiles dans la partie négative et 6 quantiles dans la partie positive. Enfin, on ajoute zéro aux valeurs obtenues.

Un réseau de neurones peut être quantifié avec différentes granularités. La pire méthode consiste à quantifier l'ensemble du réseau d'un coup. Dans ce cas, une constante commune $S$ sera obtenue pour l'ensemble du modèle. Le résultat de telles manipulations risque d'être insatisfaisant. Il est également possible de quantifier les tenseurs individuellement. Ainsi, chaque tenseur aura ses propres constantes, ou même de quantifier des lignes ou des colonnes de chaque tenseur. Ainsi, chaque ligne ou colonne aura sa propre constante. Il est également possible de diviser les tenseurs en petits blocs et d'effectuer la quantification au sein de ces blocs. Bien que les vecteurs quantifiés doivent être stockés de manière optimale, les calculs seront plus précis. En résumé, une granularité plus faible implique moins de constantes à stocker, et inversement : plus la granularité est élevée, plus les résultats des calculs quantifiés sont proches des résultats initiaux.
\section{Ajustement efficace des LLM quantifiés (QLoRA)}

Un aspect crucial de l'adaptation de bas rang (LoRA), méthode évoquée précédemment, réside dans ses besoins en mémoire pendant l'apprentissage, notamment en termes de nombre et de taille des adaptateurs utilisés. Grâce à l'empreinte mémoire minimale de LoRA, davantage d'adaptateurs peuvent être utilisés pour optimiser les performances sans augmenter significativement l'utilisation globale de la mémoire. Par conséquent, la consommation de mémoire principale lors du réglage fin de modèles de langage volumineux provient des gradients d'activation des pondérations figées du modèle d'origine, plutôt que des paramètres LoRA. Par exemple, dans un modèle LLaMA \cite{21} de 7 milliards de paramètres, entraîné sur le jeu de données FLAN v2 \cite{22} avec une taille de lot de 1, l'utilisation de la mémoire est la suivante : les pondérations LoRA, représentant généralement environ 0,2 

Cependant, il ne suffit pas de quantifier les pondérations du modèle original. En effet, lors de l'addition des résultats des adaptateurs LoRA et des pondérations du modèle, leur type doit être identique. De plus, il est inefficace d'effectuer un apprentissage sur des adaptateurs sous une forme quantifiée (int4 ou int8). À cet égard, les auteurs ont développé une nouvelle approche QLoRA \cite{16}, qui sera abordée plus loin. À l'aide des composants décrits ci-dessus, nous définissons QLoRA pour une couche linéaire unique dans le modèle de base quantifié avec un seul adaptateur LoRA comme suit :
\[
\text{\textbf{Y}}^\text{BF16} = \text{\textbf{X}}^\text{BF16}\text{dequant(\textbf{W}}^\text{NF4}) + \text{\textbf{X}}^\text{BF16}\text{\textbf{A}}^\text{BF16}\text{\textbf{B}}^\text{BF16},
\]
où $A$ et $B$ sont les adaptateurs LoRA et $W$ la matrice quantifiée de la couche linéaire initiale. Pour la mise à jour des paramètres, seul le gradient par rapport à la fonction d'erreur $L$ est nécessaire pour les pondérations d'adaptateur $\frac{\partial L}{\partial A}$ et $\frac{\partial L}{\partial B}$, et non pour les pondérations 4 bits $\frac{\partial L}{\partial W}$. Cependant, le calcul de $\frac{\partial L}{\partial A}$ et $\frac{\partial L}{\partial B}$ implique le calcul de $\frac{\partial X}{\partial W}$, qui s'effectue via l'équation ci-dessus avec déquantification du stockage $\text{\textbf{W}}^\text{NF4}$ vers le type de données de calcul $\text{\textbf{W}}^\text{BF16}$ pour calculer la dérivée $\frac{\partial X}{\partial W}$ en précision BFloat16.

Au lieu du type NF4, il peut y avoir d'autres types de quantification, ainsi que la double quantification : une méthode dans laquelle le nombre de bits diminue progressivement, en appliquant séquentiellement deux quantifications avec une taille de bloc croissante, selon laquelle la quantification est effectuée.\section{Résumé texte à texte}

Le résumé de texte est une tâche essentielle du traitement automatique du langage naturel (TALN). Il consiste à générer des résumés concis et cohérents de textes volumineux. L'objectif est de capturer les informations les plus importantes et de les transmettre sous une forme concise, tout en préservant le sens et le contexte d'origine. Le résumé de texte présente un large éventail d'applications et de défis. Il est utilisé dans les résumés d'actualités pour créer des résumés concis d'articles de presse, dans les résumés de documents pour résumer de longs rapports, articles de recherche et documents juridiques, et dans les résumés de dialogues pour condenser des conversations ou des transcriptions de réunions. Cependant, il présente également plusieurs défis, tels que le maintien de la cohérence pour garantir que le résumé soit logiquement cohérent et lisible, la couverture de toutes les informations critiques tout en restant concis, la gestion des ambiguïtés pour garantir que le résumé généré reflète fidèlement le texte d'origine, et la création de métriques efficaces pour évaluer la qualité et l'exactitude des résumés. Il existe deux principales approches du résumé de texte : le résumé extractif et le résumé abstractif.

Le résumé extractif \cite{25, 23, 24,  amini:hal-01335857} consiste à sélectionner et à extraire des phrases, expressions ou sections clés directement du texte original. Cette approche identifie les parties les plus importantes du texte selon divers critères tels que l'importance des phrases, la fréquence des mots et leur importance positionnelle. Le contenu sélectionné est ensuite combiné pour former le résumé. Il s'appuie sur la sélection de segments existants du texte, ce qui simplifie la création des résumés, car ils sont composés de phrases réelles du texte original. L'utilisation des mots et expressions exacts de la source évite les erreurs de reformulation ou la perte d'exactitude factuelle. Les techniques utilisées en résumé extractif incluent les méthodes basées sur la fréquence \cite{26} qui identifient et extraient les phrases présentant la fréquence la plus élevée de mots importants, les méthodes basées sur des graphes qui utilisent des algorithmes comme PageRank \cite{27} pour noter et classer les phrases en fonction de leurs relations et de leur importance au sein du texte, et les approches d'apprentissage automatique qui utilisent des techniques d'apprentissage supervisé ou non supervisé pour identifier et sélectionner les phrases clés.

En revanche, le résumé abstractif \cite{7, 8, 28} consiste à générer de nouvelles phrases qui restituent l'essence du texte original. Cette méthode va au-delà de la simple sélection de phrases : elle paraphrase et synthétise l'information pour créer un résumé cohérent et concis. Le résumé abstractif implique la génération d'un nouveau texte qui peut ne pas apparaître directement dans le document original. Plus complexe que le résumé extractif, il nécessite de comprendre et de reformuler le contenu. Cette approche permet de produire des résumés plus lisibles et plus réalistes, en reformulant et en condensant l'information de manière plus naturelle et cohérente. Parmi les techniques utilisées pour le résumé abstractif, on trouve les modèles Seq2Seq, qui utilisent des architectures encodeur-décodeur où l'encodeur traite le texte d'entrée et le décodeur génère le résumé ; les modèles Transformer, qui exploitent des modèles comme BERT \cite{29} ou GPT pour comprendre le contexte et générer des résumés ; et l'apprentissage par renforcement, qui permet d'affiner les modèles en récompensant les résumés cohérents et précis. Un résumé abstractif peut aider à saisir les sentiments nuancés exprimés dans les tweets, qui sont souvent concis et informels et aider à améliorer des tâches comme la classification de sentiments \cite{balikas-amini-2016-twise}. Le résumé abstractif peut aussi générer des extraits plus cohérents et informatifs pour les résultats de recherche, améliorant ainsi l'expérience utilisateur en fournissant des informations concises et pertinentes \cite{10.1007/978-3-031-63536-6_20}. Les agrégateurs de nouvelles peuvent utiliser le résumé abstractif pour fournir des résumés brefs et cohérents d'articles d'actualité, ce qui permet aux utilisateurs de parcourir plus facilement les titres et d'obtenir rapidement l'essentiel de l'actualité \cite{Alaa20}.

\subsection{MBart}

Le modèle MBart \cite{7}, ou BART multilingue, est un puissant modèle séquence à séquence basé sur des transformateurs, conçu pour la traduction automatique multilingue et la génération de texte. Développé par Facebook AI, MBart étend les capacités du modèle BART (Transformateurs bidirectionnels et autorégressifs) original à la prise en charge de plusieurs langues, lui permettant ainsi de gérer un large éventail de paires de langues et de scénarios multilingues avec une efficacité et une précision élevées.

MBart s'appuie sur une approche de pré-apprentissage et d'ajustement fin. Durant la phase de pré-apprentissage, le modèle est entraîné sur un vaste corpus de données multilingues à l'aide d'un auto-encodeur de débruitage. Cela implique de corrompre le texte d'entrée en masquant des segments de jetons, puis d'entraîner le modèle à reconstruire le texte original. Ce processus permet à MBart d'apprendre des représentations linguistiques robustes et de comprendre les nuances syntaxiques et sémantiques de différentes langues.

L'une des principales caractéristiques de MBart est sa capacité à effectuer des traductions « zero-shot », ce qui signifie qu'il peut traduire entre des paires de langues jamais vues lors de l'apprentissage. Pour ce faire, des jetons spécifiques à chaque langue aident le modèle à identifier les langues source et cible, lui permettant ainsi de généraliser efficacement entre les langues. Par exemple, lors d'une traduction de l'anglais vers le français, des jetons spécifiques indiquant la langue source (anglais) et la langue cible (français) sont ajoutés à la séquence d'entrée, guidant le modèle pour générer la traduction appropriée.

L'architecture de MBart repose sur un framework encodeur-décodeur, où l'encodeur et le décodeur sont composés de plusieurs couches de transformateurs. L'encodeur traite le texte d'entrée et génère une série d'intégrations contextuelles, tandis que le décodeur utilise ces intégrations pour produire le texte traduit ou généré, un jeton à la fois. Cette conception permet à MBart de capturer les dépendances à longue portée et de générer des sorties cohérentes et contextuellement adaptées.

MBart prend en charge un large éventail de langues, ce qui le rend polyvalent pour les applications multilingues. Il est particulièrement efficace dans les situations linguistiques à faibles ressources, où il peut exploiter les représentations partagées de langues à ressources élevées pour améliorer la qualité de la traduction. Sa capacité à gérer diverses langues et son pré-apprentissage sur des données multilingues complètes en font un outil robuste pour des tâches telles que la traduction de documents, la génération de textes multilingues et la compréhension interlinguistique.

\subsection{T5}

Le modèle T5 \cite{8}, également appelé « Transformateur de transfert de texte à texte », est un modèle polyvalent et puissant basé sur un transformateur, développé par Google Research. T5 est conçu pour gérer un large éventail de tâches de traitement automatique du langage naturel (TALN) en les présentant toutes comme des problèmes de texte à texte. Cela signifie que l'entrée et la sortie sont toujours traitées comme des chaînes de texte, ce qui simplifie l'architecture du modèle et le processus d'apprentissage.

T5 repose sur l'idée qu'un modèle unique peut effectuer diverses tâches si celles-ci sont représentées dans un format cohérent. Par exemple, des tâches telles que la traduction, le résumé, la réponse à des questions et la classification sont toutes converties en formats texte à texte. Par exemple, la traduction d'une phrase de l'anglais vers le français est formatée comme suit : « traduire l'anglais vers le français : [phrase d'entrée] », tandis que le résumé d'un paragraphe est formaté comme suit : « résumer : [texte d'entrée] ».

L'architecture de T5 repose sur le modèle de transformateur, exploitant spécifiquement la structure encodeur-décodeur. L'encodeur traite le texte d'entrée et génère une série d'incorporations contextuelles, que le décodeur utilise ensuite pour produire le texte de sortie souhaité. Cette conception permet à T5 de capturer les dépendances complexes au sein du texte et de générer des sorties de haute qualité et contextuellement pertinentes.

L'un des principaux atouts de T5 réside dans son pré-entraînement approfondi sur un ensemble de données volumineux et diversifié appelé C4 (Colossal Clean Crawled Corpus) \cite{30}. Cet ensemble de données comprend un grand nombre de pages web, offrant au modèle une compréhension approfondie du langage. Lors du pré-entraînement, T5 est exposé à diverses tâches, apprenant à prédire les jetons manquants et à comprendre différentes structures linguistiques. Ce pré-entraînement approfondi confère à T5 une solide compréhension générale, qu'il peut appliquer à un large éventail de tâches en aval, moyennant des ajustements précis.

En termes de performances, T5 a obtenu des résultats de pointe sur de nombreux benchmarks de TALN, notamment les ensembles de données GLUE, SuperGLUE et SQuAD. Sa capacité à gérer des tâches diverses avec une approche unifiée simplifie le déploiement des systèmes de TALN et facilite leur adaptation aux nouvelles tâches en modifiant simplement le préfixe d'entrée et en affinant les données pertinentes.

En résumé, le modèle T5 est un outil innovant et flexible dans le domaine du traitement du langage naturel. En traitant chaque tâche comme un problème de conversion texte-texte, T5 simplifie l'architecture du modèle et le processus d'apprentissage, lui permettant d'exceller dans un large éventail de tâches. Sa structure encodeur-décodeur, son pré-apprentissage approfondi sur le corpus C4 et l'utilisation de préfixes de tâches font de T5 un modèle puissant et polyvalent pour une multitude d'applications de TALN.

\subsection{Métriques de résumé}

Plusieurs métriques permettent de comparer la précision d'un résumé à celle d'un résumé ciblé. Il est évident qu'un bon résumé ne doit pas nécessairement correspondre exactement à celui ciblé ; les métriques utilisées doivent donc en tenir compte. Par exemple, la réorganisation de certains mots dans la valeur cible ne doit pas conduire à une valeur nulle, tandis qu'une comparaison mot à mot aboutira exactement à ce résultat.

L'un des métriques les plus populaires est la famille ROUGE \cite{31}. ROUGE est un ensemble de métriques qui comparent le chevauchement entre le résumé généré et un résumé de référence. Ces scores vont de 0 à 1, les valeurs les plus élevées indiquant une meilleure correspondance entre le résumé automatisé et la référence. Les variantes les plus courantes sont ROUGE-N, ROUGE-L et ROUGE-S. ROUGE-N mesure le chevauchement des $n$-grammes entre le résumé généré et le résumé de référence :
\[
\text{ROUGE-N} = \frac{ \sum_{n\text{-gramme} \in S} \text{Count}_{\text{correspondance}}(n\text{-gramme})}{\sum_{n\text{-gramme} \in S} \text{Count}(n\text{-gramme})},
\]
où $n$-gramme est une séquence contiguë de $n$ éléments du texte, $\text{Count}_{\text{correspondance}}(n\text{-gramme})$ est le nombre de $n$-grammes co-occurrents dans les résumés généré et de référence, et $\text{Count}(n\text{-gramme})$ est le nombre total de $n$-grammes dans le résumé de référence. Plus précisément, la valeur de $\text{Count}_{\text{match}}(n\text{-gram})$ est le minimum parmi les $n\text{-gram}$ du résumé ciblé et généré (noté $S$ dans la formule ci-dessus). Lorsqu'il existe plusieurs résumés de référence, les scores ROUGE individuels sont calculés par référence et la moyenne est renvoyée.

ROUGE-L utilise la plus longue sous-séquence commune (LCS) entre le résumé généré et le résumé de référence :
\[
\text{ROUGE-L} = \frac{LCS(X, Y)}{\text{length}(Y)},
\]
Où $LCS(X,Y)$ est la longueur de la plus longue sous-séquence commune entre les séquences $X$ (résumé généré) et $Y$ (résumé de référence). Cette métrique varie également de 0 à 1.

ROUGE-S, également appelé ROUGE-Skip-Bigram, mesure le chevauchement des skip-bigrammes entre un résumé généré et un résumé de référence. Un skip-bigramme correspond à n'importe quelle paire de mots dans l'ordre de la phrase, en tenant compte des espaces libres arbitraires. Cette métrique est particulièrement utile pour détecter la présence de paires de mots importantes qui peuvent ne pas apparaître consécutivement, mais qui conservent néanmoins leur ordre et leur pertinence contextuelle. La formule de ROUGE-S est :
\[
\text{ROUGE-S} = \frac{\sum_{\text{skip-bigram} \in S} \text{Count}_{\text{match}}(\text{skip-bigram})}{ \sum_{\text{skip-bigram} \in S} \text{Count}(\text{skip-bigram})},
\]
où skip-bigram représente les paires de mots dans leur ordre d'origine, en tenant compte des espaces vides, et $\text{Count}_{\text{match}}()$, $\text{Count}()$ sont définis comme dans la définition de ROUGE-N. ROUGE-S conserve les paires de mots importantes dans leur ordre d'origine, ce qui contribue à préserver l'intégrité contextuelle et sémantique du résumé. Il est particulièrement utile lorsque l'ordre exact des mots est moins important que la compréhension des relations entre les termes clés.

Il existe d’autres mesures, mais dans cette étude nous nous concentrerons sur cette famille de mesures.

\subsection{Ensemble de données pour le résumé de textes}

Il existe actuellement de nombreux ensembles de données pour le résumé de textes. Cependant, la plupart sont de petite taille ou très spécialisés. Cela s'explique par la difficulté de créer un résumé pour un grand nombre de textes, car cela nécessite beaucoup de ressources humaines. De plus, il est important que l'ensemble de données d'entraînement contienne des contenus courts et abstractifs, et non des contenus extractifs. Dans le cas contraire, le modèle ne sera pas entraîné pour ce type de tâche et collectera un résumé des phrases complètes du texte original.

L'actualité est l'une des principales sources de résumé de textes abstractifs. Très souvent, les textes consacrés à divers événements sont accompagnés d'un court résumé afin d'augmenter le nombre de vues. Ces textes sont rédigés par les auteurs eux-mêmes et répondent donc aux critères décrits ci-dessus. De plus, la collecte de tels textes respecte tous les droits d'auteur, car les sites d'actualités permettent l'utilisation de textes à des fins non commerciales pour les citer et les utiliser pour d'autres travaux. L'ensemble de données Gazeta proposé par Ilya Gusev \cite{32} a été sélectionné parmi une classe d'ensembles de données similaires.

Gazeta est un journal d'actualités en langue russe qui diffuse à la fois des informations locales et internationales. L'ensemble de données ne contient que les résumés de moins de 85 mots et de plus de 15 mots, ainsi que les textes de moins de 1 500 mots, et les paires texte-résumé présentant une intersection unigramme supérieure à 30 

\begin{table}[htbp!]
\centering
\begin{tabular}{||c | c | c | c ||}
\hline
& Train & Val & Test \\
\hline
1-grammes & 34,2 & 30,5 & 30,6\\
\hline
1-grammes lemmatisés & 21,4 & 17,8 & 17,6 \\
\hline
2-grammes & 68,6 & 65,0 & 65,5 \\
\hline
2-grammes lemmatisés & 61,4 & 58,0 & 58,5 \\
\hline
3-grammes & 84,5 & 81,5 & 81,9 \\
\hline
\end{tabular}
\caption{Moyenne \% de nouveaux $n$-grammes dans l'ensemble de données Gazeta}
\label{tb:1}
\end{table}
En plus de L'étude a pris en compte un ensemble de données en langue russe, avec une approche différente pour la création de contenu court dans l'ensemble de données d'entraînement. Le principal problème des ensembles de données de résumé existants, constitués d'articles de presse, est le suivant : ces articles sont rédigés par des journalistes et respectent le style journalistique. En tant que rédacteurs professionnels, les journalistes hiérarchisent et structurent généralement leurs textes en commençant par mentionner les éléments les plus importants et les plus accrocheurs d'un article dans les paragraphes d'introduction, puis en ajoutant des détails et des informations contextuelles. Ce style d'écriture pourrait expliquer le faible pourcentage de résumés à un et deux grammaires dans l'ensemble de données Gazeta (les résumés sont généralement formés en répétant quelques premières phrases du texte ou en les modifiant légèrement).

Ils obtiennent généralement de meilleurs résultats que les systèmes de résumé existants. Les auteurs ont proposé une autre méthode de génération de résumés \cite{33}, en utilisant WikiHow, une citation avec des instructions. Cet ensemble de données contient des articles rédigés par des personnes ordinaires, et non par des journalistes. Par conséquent, les textes ne sont pas rédigés par des professionnels, ce qui est plus proche de la réalité. La base de connaissances WikiHow est un recueil d'articles en ligne proposant des instructions étape par étape sur un large éventail de sujets, des arts et du divertissement à l'informatique et à l'électronique. Chaque article comporte un titre commençant par « Comment faire » et une brève description. Les articles se répartissent en deux catégories : les articles à méthode unique, qui décrivent les tâches par étapes séquentielles, et les articles à méthodes multiples, qui proposent différentes approches pour réaliser une tâche. Chaque étape des articles commence par un résumé en gras, suivi d'une explication plus détaillée. Pour générer les résumés, seuls les articles comportant plusieurs étapes ont été retenus. Dans chaque article, les premières phrases de chaque étape ont été extraites et concaténées dans un texte distinct, considéré comme un résumé ciblé, tandis que les phrases extraites ont été retirées du texte original. L'ensemble de données final est composé de 204 004 paires d'articles et de leurs résumés.

Concernant les caractéristiques de cet ensemble de données, il est important de préciser que la longueur moyenne des articles est de 579,8 et celle du résumé de 62,1. La différence entre la longueur du résumé et celle de l'ensemble de données précédent n'étant pas significative, une comparaison plus approfondie des mesures est pertinente. De plus, un tableau présentant les pourcentages de nouveaux $n$-grammes dans les résumés par rapport au texte original, comme pour l'ensemble de données Gazeta, est disponible dans le tableau \ref{tb:2}.

\begin{table}[htbp!]
\centering
\begin{tabular}{||c | c ||}
\hline
& Ensemble de données \\
\hline
1-grammes & 31,1\\
\hline
1-grammes lemmatisés & 21,6 \\
\hline
2-grammes & 78,8 \\
\hline
2-grammes lemmatisés & 70,1 \\
\hline
3-grammes & 93,3 \\
\hline
\end{tabular}
\caption{Moyenne \% de nouveaux $n$-grammes dans l'ensemble de données WikiHow}
\label{tb:2}
\end{table}

Il est important de noter qu'en raison du fait que l'ensemble de données n'a pas été divisé en ensembles de données d'apprentissage, de validation et de test, les métriques ont été calculées pour l'ensemble de données. De plus, on constate que pour les 1-grammes, la valeur métrique est approximativement la même, tandis que le pourcentage de nouveaux 2-grammes et 3-grammes pour l'ensemble de données wikiHow est significativement plus élevé que pour l'ensemble de données Gazeta. Cette différence s'explique précisément par le fait que les articles sont rédigés par des journalistes professionnels, ce qui peut entraîner davantage de plagiat et de résumés extractifs. Cependant, il convient de noter qu'une telle comparaison n'est pas totalement exacte, car la langue des ensembles de données est différente ; par conséquent, cette statistique (pourcentage de $n$-grammes) peut fortement dépendre de la langue.

\subsection{Ajustement fin des modèles de résumé de texte}

Cette section décrit l'ajustement fin des modèles de résumé de texte T5 et MBart présentés précédemment. Après réception des résultats, j'ai analysé l'effet de l'application de LoRA et d'AdaLoRA (voir section 2) aux matrices des couches d'auto-attention. Des expériences ont été menées pour différents hyperparamètres $r$, ce qui a permis d'identifier le paramètre optimal en termes de précision et de nombre de paramètres du modèle entraîné.

\begin{figure}[htbp!]
\begin{center}
\includegraphics[width=1\columnwidth]{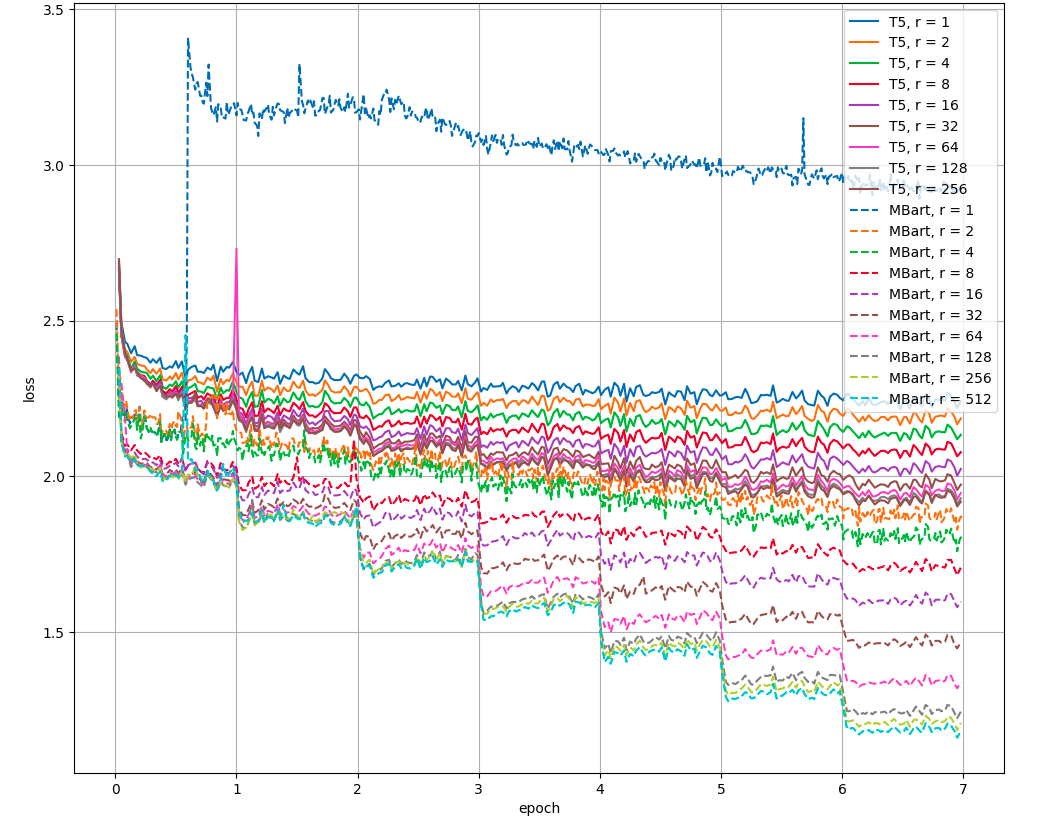}
\end{center}
\caption{Courbe d'apprentissage pour l'ajustement fin des modèles MBart et T5 avec LoRA et différents $r$}
\label{fig:training}
\end{figure}

Par exemple, le tracé d'apprentissage des modèles MBart et T5 est présenté dans la figure \ref{fig:training}.

Il convient tout d'abord de noter que plus la valeur de $r$ augmente, plus les courbes d'apprentissage diminuent. Cela signifie que la prédiction est très proche en termes de fonction de perte, qui est influencée par l'entropie croisée sur les plongements. Il est important de noter qu'une diminution de la fonction de perte ne signifie pas une amélioration du résumé obtenu, car le résumé obtenu peut différer considérablement de la cible en termes de plongement, tout en étant relativement précis. Pour évaluer la qualité du modèle, la métrique ROUGE est nécessaire, ce qui permet de visualiser plus précisément la coïncidence sémantique entre deux textes. Idéalement, l'évaluation la plus précise est une évaluation humaine. Malheureusement, il est impossible d'évaluer la qualité du modèle après chaque étape du gradient, car le calcul de la métrique ROUGE sur la puissance de calcul du cluster prend en moyenne 40 minutes. À cet égard, le nombre d'époques nécessaires à l'apprentissage a été estimé en fonction du degré de diminution de la forêt, puis la métrique ROUGE a été calculée sur les modèles finaux (voir les tableaux \ref{tb:3} et \ref{tb:4}). Le graphique de la figure \ref{fig:training} montre que pour la plupart des modèles, 7 époques suffisent à l'apprentissage. Bien qu'il soit clair que pour les modèles avec un hyperparamètre important, $r$ peut être entraîné davantage, puisque la tendance à la baisse de la fonction de perte continue de diminuer, il a été vérifié que cela n'a pas d'effet significatif sur la métrique ROUGE. Concernant le nombre d'époques requises pour l'entraînement des modèles, il est important de noter que le nombre d'époques importe peu, mais plutôt le nombre d'étapes (c'est-à-dire le calcul du gradient sur le lot) effectuées pendant l'entraînement. C'est pourquoi l'entraînement du jeu de données WikiHow a nécessité moins d'époques, cinq au lieu de sept. Cependant, le processus d'achèvement a été le même en termes de temps, car l'échantillon d'entraînement était plus grand. Par conséquent, le nombre total d'étapes pour les jeux de données Gazeta et WikiHow était approximativement identique. Enfin, notons que le temps d'entraînement d'une époque varie de 40 à 100 minutes, selon le nombre de paramètres (c'est-à-dire la valeur du paramètre $r$) pour les adaptateurs LoRA.

Concentrons-nous maintenant sur la valeur de la métrique ROUGE pour les expériences réalisées. Comme mentionné précédemment, un entraînement similaire à celui dont le graphique est présenté ci-dessus a été réalisé sur deux jeux de données pour les modèles T5 et MBart, avec des pondérations entièrement ajustées, ainsi qu'un ajustement fin avec des adaptateurs LoRA et AdaLoRA. Les résultats sont présentés dans les tableaux \ref{tb:3} et \ref{tb:4}. Dans les tableaux concernant les cas où des adaptateurs LoRA ont été utilisés, les meilleures valeurs sont mises en évidence pour tous les $r$ possibles, afin de ne pas surcharger le tableau. La dépendance des métriques à $r$ sera abordée ultérieurement.

\begin{table}[htbp!]
\centering
    \resizebox{\textwidth}{!}{
\begin{tabular}{||c | c | c | c | c ||}
\hline
& ROUGE-1 & ROUGE-2 & ROUGE-L & ROUGE-S \\
\hline
MBart ajusté & 32,1 & 14,2 & \textbf{27,9} & 20,1 \\
\hline
T5 ajusté & \textbf{35,3} & 13,1 & 26,5 & 22,4 \\
\hline
MBart + LoRA (meilleur) & 24,1 & 9,7 & 18,9 & 18,9 \\
\hline
T5 + LoRA (meilleur) & 24,5 & 9,9 & 19,3 & 19,3 \\
\hline
MBart + AdaLoRA (meilleur) & 33,4 & \textbf{17,1} & 25,9 & \textbf{23.3} \\
\hline
T5 + AdaLoRA (meilleur) & 30.1 & 15.7 & 25.4 & 20.3 \\
\hline
\end{tabular}}
\caption{Métriques ROUGE pour les modèles affinés sur l'ensemble de données Gazeta}
\label{tb:3}
\end{table}

\begin{table}[htbp!]
\centering
    \resizebox{\textwidth}{!}{
\begin{tabular}{||c | c | c | c | c ||}
\hline
& ROUGE-1 & ROUGE-2 & ROUGE-L & ROUGE-S \\
\hline
MBart ajusté & \textbf{35.9} & \textbf{13.9} & \textbf{34.8} & 25.4 \\
\hline
T5 ajusté & 35.4 & 9.3 & 23.6 & 28.4 \\
\hline
MBart + LoRA (meilleur) & 27.1 & 10.2 & 22.3 & 20.1 \\
\hline
T5 + LoRA (meilleur) & 28.5 & 9.1 & 23.3 & 26.3 \\
\hline
MBart + AdaLoRA (meilleur) & 33.4 & 13.0 & 33.1 & 26.1 \\
\hline
T5 + AdaLoRA (meilleur) & \textbf{36.1} & 12.1 & 25.4 & \textbf{29.3} \\
\hline
\end{tabular}}
\caption{Métriques ROUGE pour les modèles affinés sur le jeu de données WikiHow}
\label{tb:4}
\end{table}

Parmi les résultats obtenus, il est important de noter que dans les deux cas, les valeurs des métriques lors de l'apprentissage avec des adaptateurs se sont avérées proches de celles obtenues si le modèle avait été entièrement entraîné. En moyenne, les métriques et la qualité sont légèrement inférieures, mais le temps consacré à l'entraînement a considérablement diminué. Ainsi, il faut en moyenne environ une heure par époque pour entraîner un modèle avec des adaptateurs LoRA, contre 34 heures en moyenne pour un modèle complet. Il s'agit d'une différence considérable, sachant que le modèle doit être entraîné sur plusieurs époques pour atteindre une qualité adéquate.

Un autre résultat important démontré lors de ces expériences est la supériorité absolue de la méthode AdaLoRA sur la méthode LoRA, quel que soit le jeu de données, quel que soit le modèle et la métrique de base. Cela est principalement dû à l'algorithme, qui permet de sélectionner le rang approprié de la matrice pour chaque couche, et non de le fixer de manière fixe, comme c'est le cas avec LoRA. Ce résultat important sera pris en compte dans la section suivante et lors de la définition du modèle final. Un autre fait important est qu'AdaLoRA est supérieur à un modèle entièrement affiné sur certaines métriques. Ce point a été abordé plus en détail dans la deuxième section, mais ce phénomène peut s'expliquer par le fait que les grands réseaux, après affinement, oublient certaines informations reçues lors de l'apprentissage initial, ce qui entraîne une perte de qualité. Cela confirme l'effet décrit dans l'article original sur AdaLoRA.

\begin{figure}[htbp!]
\begin{center}
\includegraphics[width=0.95\columnwidth]{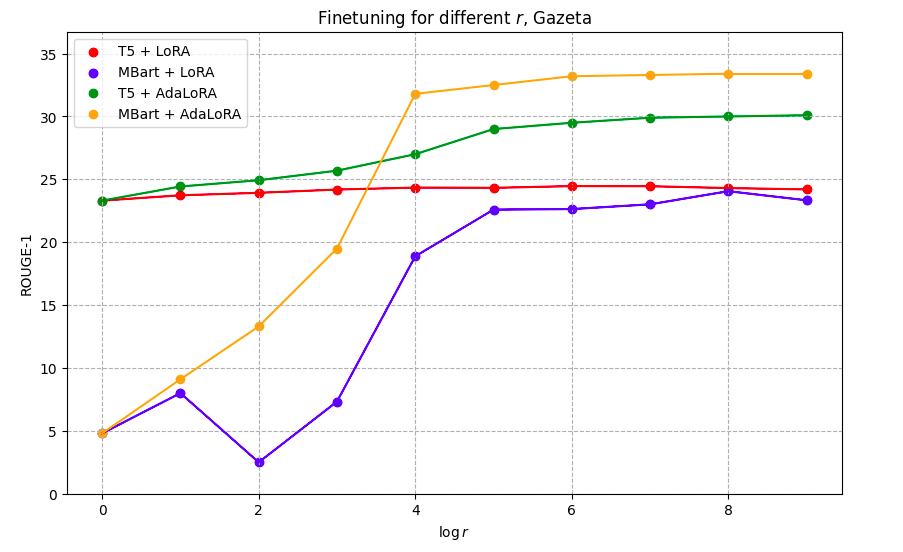}
\end{center}
\caption{Métrique ROUGE-1 pour l'ajustement fin des modèles MBart et T5 avec des adaptateurs de $r$ différents, avec le jeu de données Gazeta}
\label{fig:gazeta_rouge}
\end{figure}

\begin{figure}[htbp!]
\begin{center}
\includegraphics[width=0.95\columnwidth]{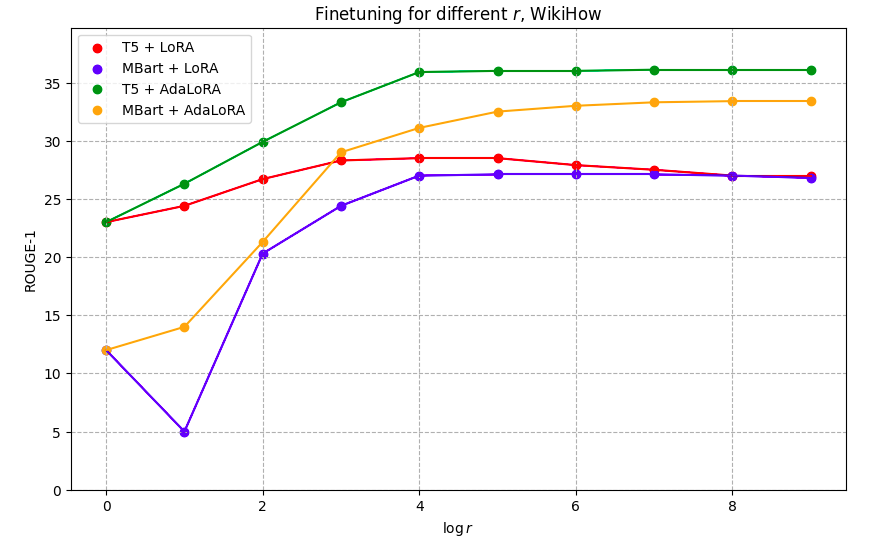}
\end{center}
\caption{Métrique ROUGE-1 pour l'ajustement fin des modèles MBart et T5 avec des adaptateurs de $r$ différents, avec WikiHow Ensemble de données}
\label{fig:wikihow_rouge}
\end{figure}

Comme annoncé précédemment, concentrons-nous maintenant sur l'évolution des métriques avec différents hyperparamètres $r$. Les graphiques \ref{fig:gazeta_rouge} et \ref{fig:wikihow_rouge} ci-dessous illustrent l'évolution de la métrique ROUGE-1 pour différents $r$. Sur l'axe des abscisses, la valeur de $r$ est représentée sur une échelle logarithmique. Pour les autres métriques, des graphiques présentant des tendances similaires sont obtenus ; ils ne sont donc pas présentés ici.

La première chose que l'on remarque sur les graphiques est que, pour chaque modèle, à faible r, les valeurs des métriques LoRA et AdaLoRA sont très proches. Cependant, lorsque la valeur de r augmente, la métrique ROUGE-1 augmente également, ce qui s'améliore pour AdaLoRA. Une autre caractéristique démontrée par ces graphiques est que la valeur optimale de la métrique est obtenue avec des valeurs suffisamment faibles de l'hyperparamètre r, approximativement dans un diapason de 16 à 32. Au-delà, la valeur seuil reste la même ou se dégrade légèrement. Enfin, on remarque que pour des valeurs élevées de r, le résultat de l'ajustement fin des modèles avec LoRA s'avère indépendant du modèle d'apprentissage. Parallèlement, pour AdaLoRA, le meilleur résultat est obtenu par le modèle qui, même après un apprentissage complet des échelles, obtient la meilleure valeur métrique.

\section{Reconnaissance automatique de la parole}

\subsection{Whisper}

Whisper \cite{9} est une famille de modèles de reconnaissance automatique de la parole (RAP) développée par OpenAI. Elle représente une avancée significative dans le domaine de la reconnaissance vocale grâce à sa robustesse et sa polyvalence. Ces modèles sont basés sur l'architecture Transformer, un modèle largement adopté dans les tâches de traitement automatique du langage naturel (TALN). Cette architecture se compose d'une structure encodeur-décodeur, où l'encodeur et le décodeur sont composés de plusieurs couches de réseaux de neurones auto-attentionnels et à anticipation. L'encodeur traite le signal audio d'entrée, en extrait les caractéristiques et crée une représentation qui capture les informations essentielles de l'audio. Le décodeur utilise ensuite cette représentation codée et génère la transcription textuelle correspondante, en prenant en compte différentes parties de l'entrée codée et en utilisant le contexte des jetons précédemment générés pour prédire le jeton suivant.

Les modèles Whisper sont disponibles en différentes tailles, chacune se distinguant par le nombre de couches et la taille des couches de l'encodeur et du décodeur. Il en résulte des modèles avec différents nombres de paramètres, conçus pour équilibrer efficacité et précision de calcul. Par exemple, Whisper-Tiny, le plus petit modèle conçu pour les applications à faible latence, compte environ 39 millions de paramètres. Whisper-Base est un modèle légèrement plus grand avec environ 74 millions de paramètres. Whisper-Small, un modèle intermédiaire, compte environ 244 millions de paramètres. Whisper-Medium est un modèle plus grand avec environ 769 millions de paramètres. Le plus grand modèle, Whisper-Large, offre la plus grande précision avec environ 1,55 milliard de paramètres. Des expériences seront menées ultérieurement avec chacun de ces modèles et les résultats des mesures seront comparés.

Ces modèles ont été entraînés à l'aide d'un ensemble de données vaste et diversifié d'enregistrements audio et de leurs transcriptions correspondantes. Cet ensemble de données comprend une grande variété de langues, d'accents et de conditions acoustiques, ce qui permet au modèle de bien se généraliser à différents contextes vocaux. Le processus d'entraînement consiste à convertir les signaux audio en spectrogrammes, qui sont des représentations temps-fréquence de l'audio servant d'entrée au modèle. Diverses techniques d'augmentation des données sont appliquées pour accroître artificiellement la diversité des données d'entraînement, comme l'ajout de bruit de fond, la variation de la hauteur tonale et la modification de la vitesse audio. Le modèle est entraîné à l'aide de techniques d'optimisation par descente de gradient, comme Adam, afin de minimiser l'erreur entre les transcriptions prédites et la réalité terrain.

Les modèles Whisper sont reconnus pour leur robustesse et leur capacité à gérer un large éventail de variations vocales, notamment entre différentes langues, accents et dialectes, grâce à la diversité de leurs données d'entraînement. Ils fonctionnent également bien dans les environnements bruyants, ce qui les rend adaptés aux applications concrètes où le bruit de fond est prédominant. Ces modèles peuvent être appliqués à de nombreuses tâches, notamment pour améliorer la précision et la réactivité des assistants vocaux tels que Siri, Alexa et Google Assistant, fournir des transcriptions précises pour les réunions, les entretiens et les contenus multimédias, proposer des services de transcription et de sous-titrage en temps réel pour les malentendants, et accompagner les apprenants en langues grâce à des transcriptions et des retours de prononciation précis. En résumé, les modèles Whisper, avec leur architecture avancée, leur formation approfondie et leurs grandes tailles de paramètres, représentent une solution de pointe dans le domaine de la reconnaissance automatique de la parole, offrant une grande précision et une grande polyvalence dans diverses applications et langues.

\subsection{Jeu de données et métrique}

Trouver un jeu de données pour l'entraînement du modèle ASR n'a pas été une tâche aisée. Bien qu'il existe de nombreux jeux de données pour l'entraînement des modèles Audio2Text, disposer de paires audio-texte ne suffisait pas ; nous avions également besoin de résumés pour chaque texte. Il existe très peu de jeux de données de ce type, et l'un d'eux est le jeu de données How2 \cite{37}. Bien que la raison de la nécessité d'un contenu concis soit abordée dans la section suivante, il est important de le souligner pour comprendre la motivation de l'utilisation de ce jeu de données particulier.

Le jeu de données How2 est une riche collection de vidéos pédagogiques, chacune accompagnée d'énoncés oraux, de sous-titres en anglais, de traductions participatives en portugais et de résumés vidéo en anglais. La multimodalité étendue de ce jeu de données en fait une excellente ressource pour le développement de modèles avancés de compréhension multimodale. Contrairement à d'autres jeux de données multimodaux, How2 utilise des données naturelles : les sous-titres et les résumés sont créés par les créateurs des vidéos originales et non par le crowdsourcing, ce qui garantit des annotations authentiques et contextuellement pertinentes. Le contenu visuel des vidéos est intrinsèquement lié aux énoncés parlés, offrant des fonctionnalités supplémentaires pour améliorer l'apprentissage du modèle. L'ensemble de données comprend 79 114 vidéos pédagogiques, totalisant 2 000 heures de contenu, d'une durée moyenne de 90 secondes. Ces vidéos proviennent de YouTube et sont accompagnées de différents types de métadonnées, notamment des sous-titres et des descriptions de référence en anglais.

Outre la fonction de perte, utilisée pour entraîner le modèle, il est nécessaire de déterminer la métrique permettant de comparer la qualité des modèles. Le taux d'erreur de mots (WER) est une métrique courante utilisée pour évaluer la précision des systèmes de reconnaissance automatique de parole (RAP). Il mesure la différence entre une transcription de référence, qui est la transcription correcte, et la transcription d'hypothèse produite par le système de reconnaissance automatique de parole. Le WER est largement utilisé car il offre un moyen simple et interprétable de quantifier les erreurs dans les résultats de reconnaissance vocale. La formule de calcul du WER est la somme du nombre de substitutions, de suppressions et d'insertions divisée par le nombre total de mots de la transcription de référence. Pour calculer le WER, un algorithme de programmation dynamique similaire à la distance de Levenshtein est utilisé pour aligner les transcriptions de référence et d'hypothèse, permettant ainsi l'identification des substitutions, des suppressions et des insertions.

Le WER est essentiel pour comparer les performances de différents systèmes de reconnaissance vocale, des valeurs de WER faibles indiquant des transcriptions plus précises. Malgré sa simplicité et son utilisation répandue, le WER présente certaines limites. Il traite toutes les erreurs de manière égale, quel que soit le contexte, ne prend pas en compte l'exactitude syntaxique ou sémantique et ne pondère pas les erreurs en fonction de la signification des mots. Néanmoins, le WER est largement utilisé pour l'évaluation comparative des systèmes de reconnaissance vocale, l'orientation des améliorations du développement de la reconnaissance vocale et la recherche universitaire. Il reste une mesure fondamentale pour comprendre et améliorer les performances des technologies de reconnaissance vocale.

\subsection{Modèle de Whisper à réglage fin}

Si, lors de l'apprentissage du modèle de résumé de texte, l'accent a été mis sur l'étude de différentes approches de réapprentissage des adaptateurs AdaLoRA, lors de l'optimisation des modèles ASR, l'objectif de ce travail a été d'étudier l'influence de la quantification, décrite dans la section 2. Cette approche présente des avantages particulièrement précieux : elle réduit non seulement le temps d'optimisation, mais aussi les coûts mémoire du modèle obtenu.

Examinons de plus près les expériences réalisées. J'ai optimisé des modèles Whisper avec différents types de quantification (int4 et int8) avec des adaptateurs LoRA, avec ou sans quantification. L'apprentissage sans quantification n'a été réalisé que pour les petits modèles (minuscules et petits) en raison des limites de la puissance de calcul du cluster. Voici les principaux hyperparamètres utilisés lors de l'apprentissage des modèles quantifiés. Ceci est très important, car la pratique a montré que les valeurs du pas d'apprentissage présentées dans l'article sont trop grandes pour un réglage fin. Il est donc nécessaire de prendre des hyperparamètres inférieurs à ceux de l'article original (ceux proposés sont 40 fois inférieurs).

\begin{table}[ht]
\centering    \resizebox{\textwidth}{!}{
\begin{tabular}{|| >{\centering\arraybackslash}m{3,6cm}| >{\centering\arraybackslash}p{4cm} | >{\centering\arraybackslash}p{4,5cm} | >{\centering\arraybackslash}p{2cm} ||}
\hline
& pas d'apprentissage (papier) & pas d'apprentissage (proposé) & taille du lot \\
\hline
Minuscule (39 M~paramètres) & $1,5\times10^{-3}$ & $3,75\times10^{-5}$ & 8 \\
\hline
Base (74 M~paramètres) & $1\times10^{-3}$ & $2,5\times10^{-5}$ & 8 \\
\hline
Petit (244 M~paramètres) & $5\times10^{-4}$ & $1,25\times10^{-5}$ & 4 \\
\hline
Moyen (769 M~paramètres) & $2,5\times10^{-4}$ & $6,25\times10^{-6}$ & 2 \\
\hline
Grand (1550M~paramètres) & $1,75\times10^{-4}$ & $4,375\times10^{-6}$ & 1 \\
\hline
\end{tabular}}
\caption{Hyperparamètres pour le réglage fin du Whisper quantifié }
\label{tb:5}
\end{table}

Un autre point important est que les données contenues dans l'ensemble de données sont prises en charge au format Kaldi. Ce format de représentation des caractéristiques sonores n'est pas adapté au modèle Whisper. Vous pouvez donc soit supprimer les premières couches de l'encodeur, soit convertir les caractéristiques en mel-spectogram à l'aide de la bibliothèque librosa. La seconde méthode, plus simple et ne nécessitant pas d'entraînement supplémentaire du modèle, a été choisie pour cette tâche.

Avant d'aborder les indicateurs de qualité, examinons le degré de compression des modèles, le nombre de paramètres supplémentaires des adaptateurs AdaLoRA présents dans le modèle et le temps nécessaire à l'entraînement d'une époque du modèle. Ces données sont présentées dans les tableaux ci-dessous. Outre la valeur absolue de la mémoire requise pour stocker les données du modèle, on peut également en déduire la quantité de mémoire nécessaire pour réentraîner complètement ces modèles. En général, pour l'entraînement complet des modèles, il faut 10 à 15 fois plus de GPU que la pondération du modèle, et pour l'inférence, environ 1,5 fois plus de mémoire que la pondération du modèle. Il est clair que ces chiffres dépendent fortement de la taille du lot, mais ces données sont indiquées en moyenne.

\begin{table}[ht]
\centering    \resizebox{\textwidth}{!}{
\begin{tabular}{|| >{\centering\arraybackslash}m{3,6cm}| >{\centering\arraybackslash}p{2,5cm} | >{\centering\arraybackslash}p{2cm} | >{\centering\arraybackslash}p{2cm} |>{\centering\arraybackslash}p{2.5cm} ||}
\hline
& mémoire (modèle initial) & mémoire (int4) & mémoire (int8) & compression AdaLoRA \\ \hline
Minuscule (39 M~paramètres) & 75 Mo & 22 Mo & 40 Mo & 1,0 \\
\hline
Base (74 M~paramètres) & 142 Mo & 42 Mo & 76 Mo & 1,7 \\
\hline
Petit (244 M~paramètres) & 466 Mo & 140 Mo & 265 Mo & 1,8 \\
\hline
Moyen (769 M~paramètres) & 1,5 Go & 460 Mo & 905 Mo & 2,0  \\
\hline
Grande taille (1550 Mo de paramètres) & 2,9 Go & 890 Mo & 1,7 Go & 2,3 \\
\hline
\end{tabular}}
\caption{Caractéristiques mémoire de Whisper}
\label{tb:6}
\end{table}

Concernant la qualité de la compression, le tableau montre qu'avec une quantification en int4, le volume total des pondérations diminue d'environ 65 

\begin{table}[ht]
\centering
    \resizebox{\textwidth}{!}{
\begin{tabular}{|| >{\centering\arraybackslash}m{3,6cm}| >{\centering\arraybackslash}p{2,5cm} | >{\centering\arraybackslash}p{2cm} | >{\centering\arraybackslash}p{2cm} |>{\centering\arraybackslash}p{2.5cm} ||}
\hline
& modèle initial & int4 & int8 \\
\hline
Minuscule (39 M~paramètres) & 126 min & 38 min & 53 min \\
\hline
Base (74 M~paramètres) & 250 min & 75 min & 133 min \\
\hline
Petit (244 M~paramètres) & --- & 150 min & 287 min \\
\hline
Moyen (769 M~paramètres) & --- & 445 min & 823 min \\
\hline
Grand (1550 M~paramètres) & --- & 870 min & 1530 min\\
\hline
\end{tabular}}
\caption{Temps de Whisper pour différentes approches de réglage fin par époque pour environ 40 h de données d'apprentissage}
\label{tb:6}
\end{table}

\begin{figure}[!tbp]
\includegraphics[width=0.49\textwidth]{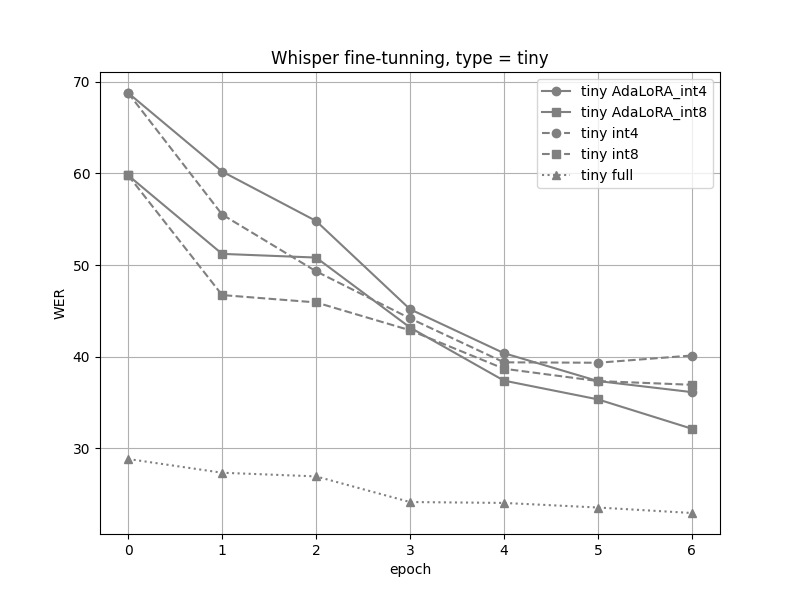} \hfill
\includegraphics[width=0.49\textwidth]{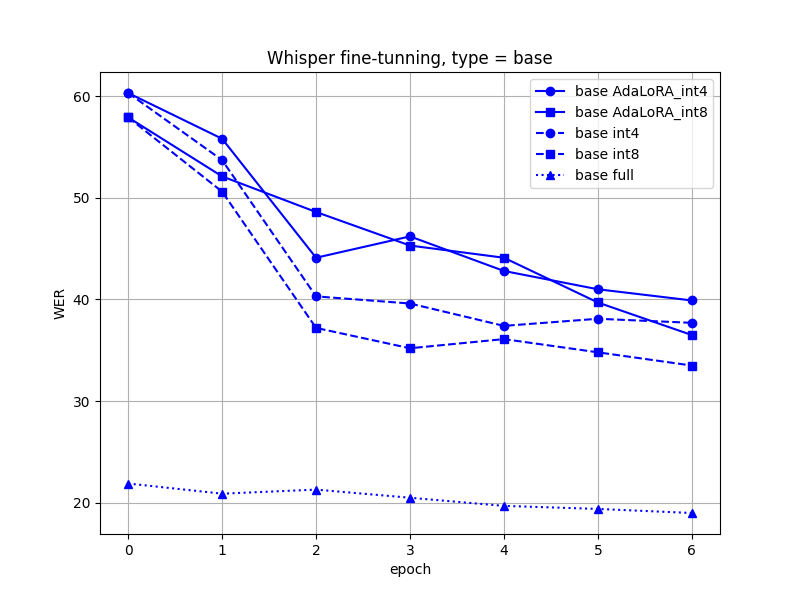}
\\[\smallskipamount]

\includegraphics[width=0.49\textwidth]{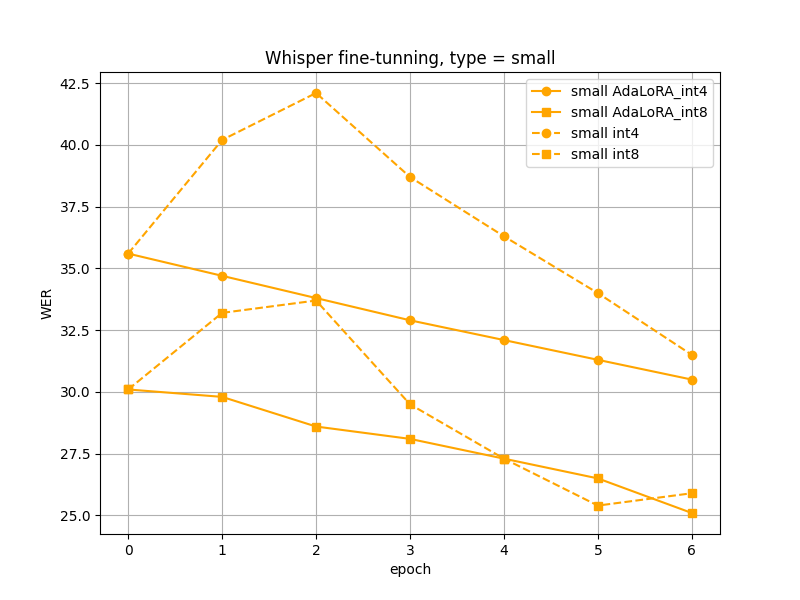}
\hfill
\includegraphics[width=0.49\textwidth]{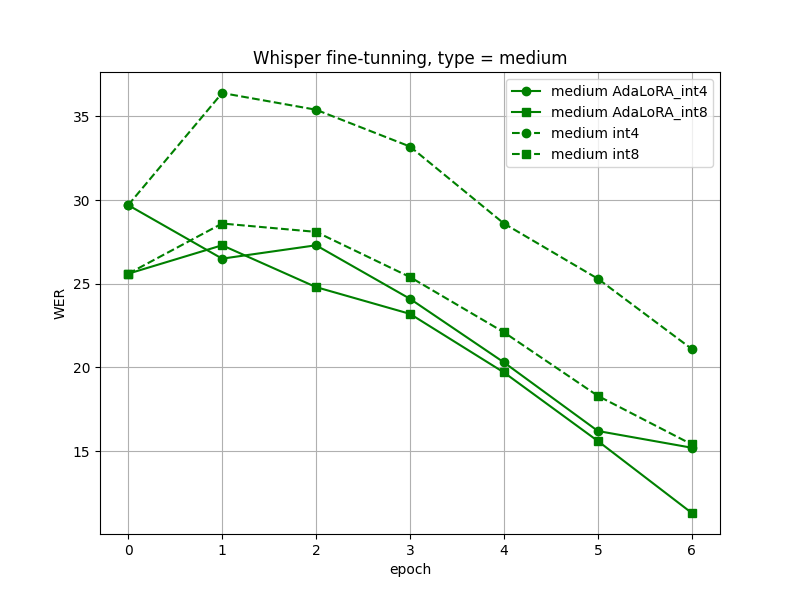}
\\[\smallskipamount]
\centering
\includegraphics[width=0.49\textwidth]{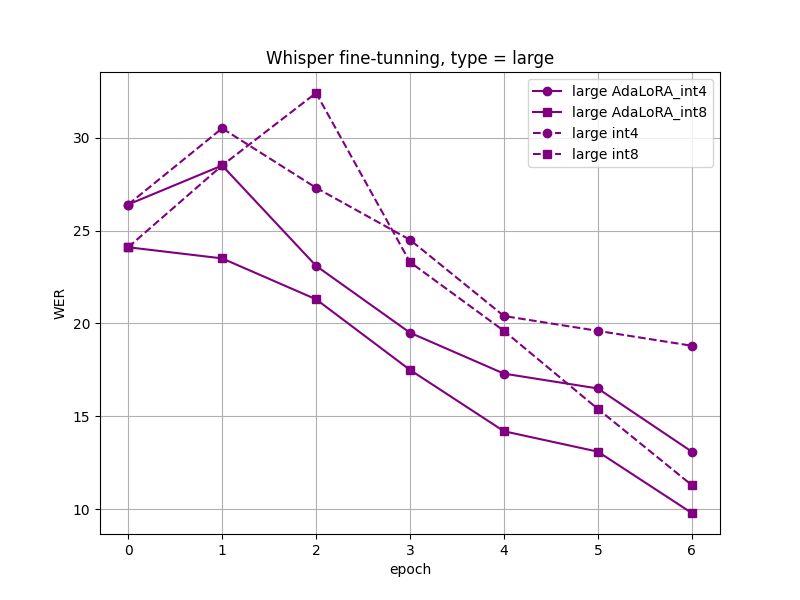}
\caption{Rendement d'erreur d'exécution (WER) lors du réglage fin du modèle Whisper avec différentes pondérations}\label{fig:whisper1}
\end{figure}

\begin{figure}[!tbp]
\includegraphics[width=0.49\textwidth]{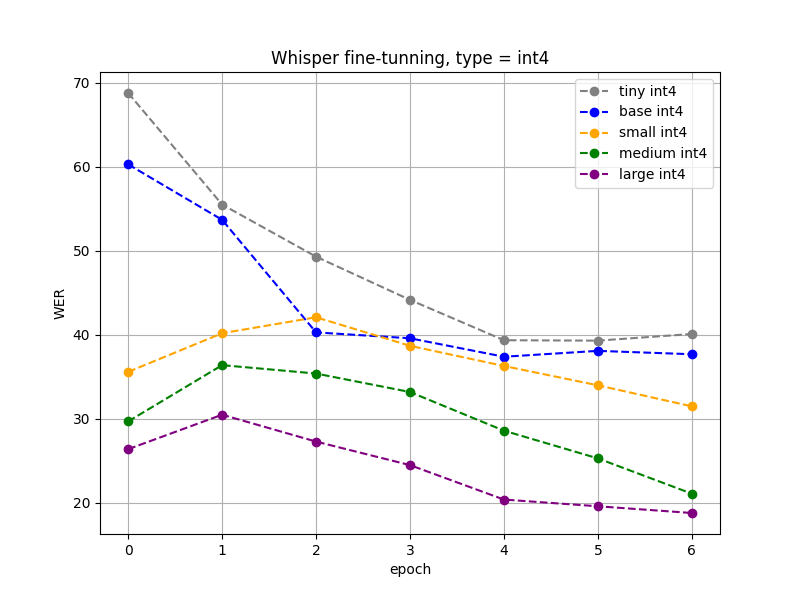} \hfill
\includegraphics[width=0.49\textwidth]{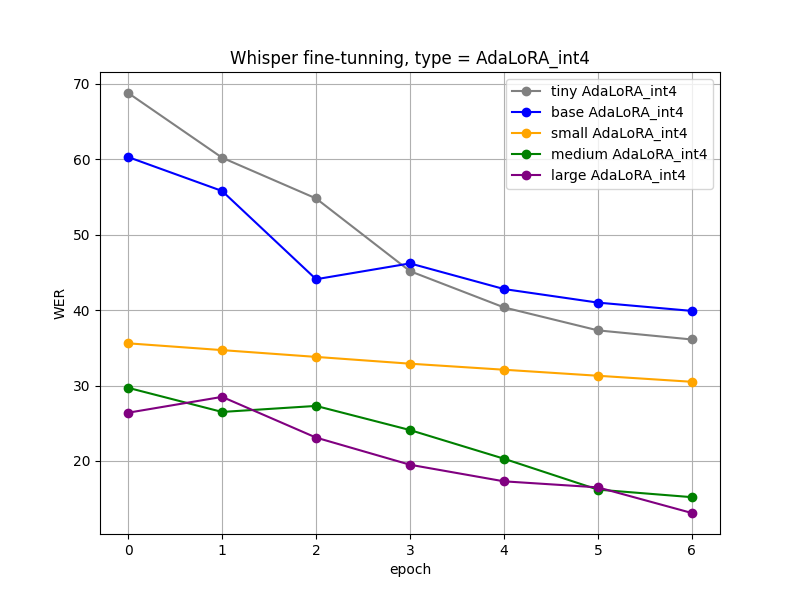}
\\[\smallskipamount]

\includegraphics[width=0.49\textwidth]{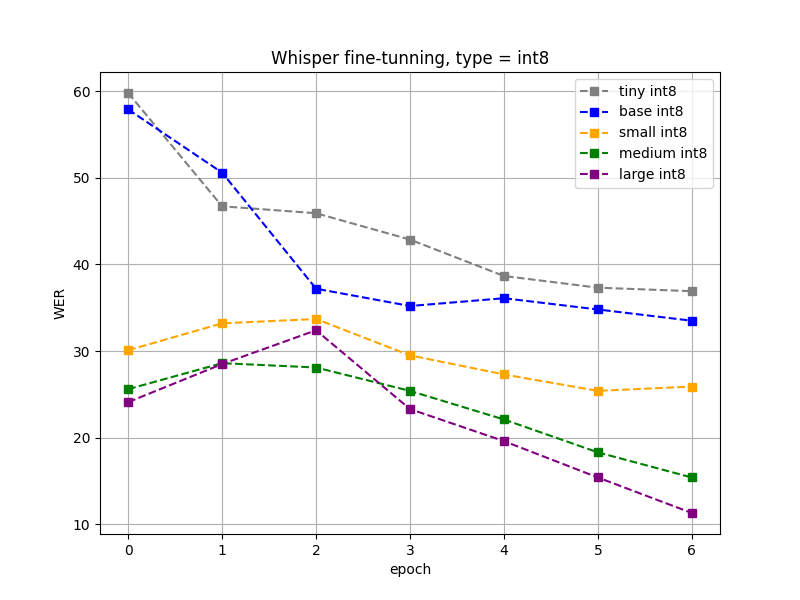}
\hfill
\includegraphics[width=0.49\textwidth]{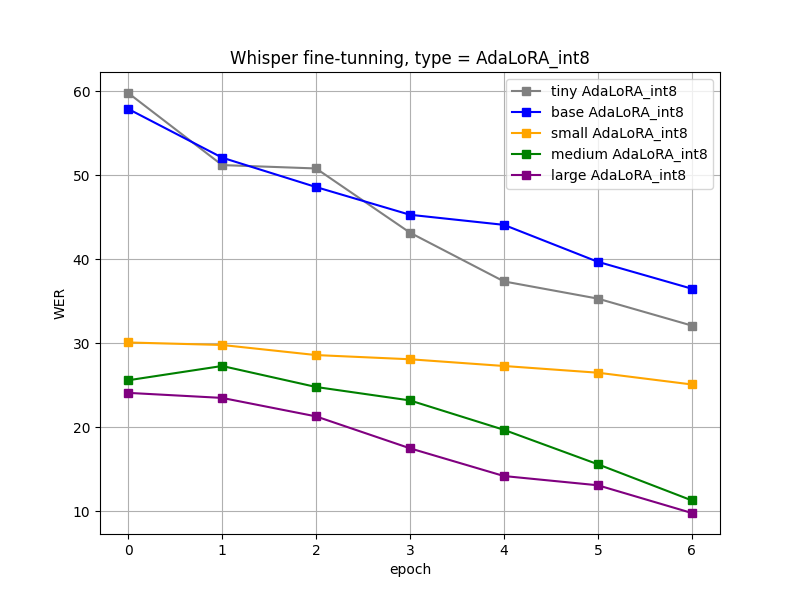}
\\[\smallskipamount]
\centering
\includegraphics[width=0.49\textwidth]{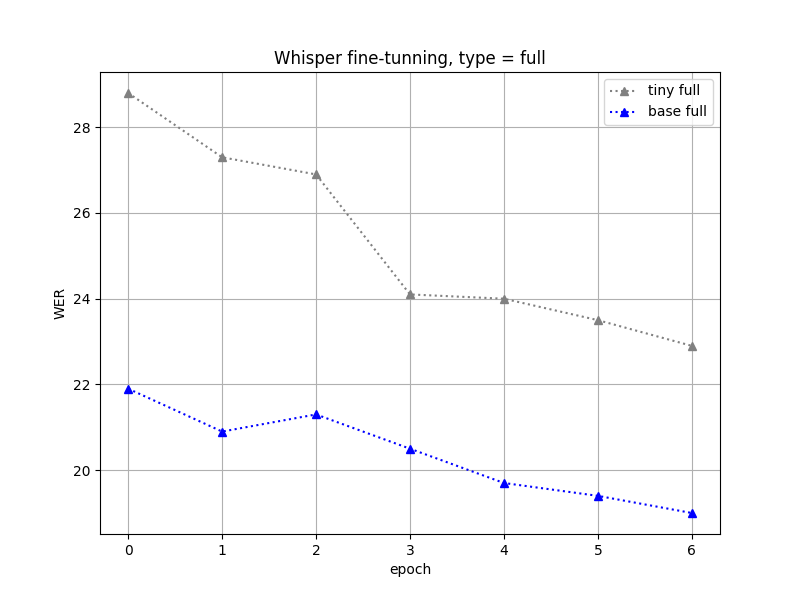}

\caption{Rendement d'apprentissage (WER) lors du réglage fin du modèle Whisper avec différents types de réglage fin }\label{fig:whisper2}
\end{figure}

Concentrons-nous maintenant sur les graphiques de la métrique WER pour différents modèles en cours d'apprentissage sur 6 époques. Vous trouverez ci-dessous 10 graphiques pour chaque type d'apprentissage (int4, int8, int4 + AdaLoRA, int8 + AdaLoRA) et chaque taille du modèle Wisper (minuscule, base, petit, moyen, grand). Ces 10 graphiques sont regroupés selon deux critères. La figure \ref{fig:whisper1} présente les graphiques des métriques décroissantes pour chacune des cinq dimensions du modèle. La figure \ref{fig:whisper2} présente les métriques pour chaque type d'apprentissage.

Les schémas suivants sont observés dans le graphique \ref{fig:whisper1}. Premièrement, on observe que les modèles minuscules et de base parviennent à apprendre en 6 époques, tandis que les modèles basés sur des réentraînements de petite, moyenne et grande taille nécessitent un entraînement supplémentaire, le nombre d'itérations de descente de gradient dans ces réseaux de neurones étant insuffisant. Une autre tendance importante est que les courbes WER des modèles minuscules et de base diminuent immédiatement dès la première époque. En revanche, ce n'est pas le cas pour les modèles de petite, moyenne et grande taille : les types d'apprentissage sans adaptateurs AdaLoRA dégradent d'abord leur métrique, puis l'augmentent. Cela pourrait s'expliquer par le fait que les grands réseaux, stockant davantage d'informations, perdent une partie des informations obtenues sur l'ensemble de données d'origine lorsque les pondérations (même quantifiées) sont entièrement affinées. De plus, un écart aussi important entre deux sous-ensembles de ces deux modèles pourrait être dû à une forte augmentation du nombre de pondérations lors du passage du modèle de base au modèle de petite taille (multiplication par 3,3). Une différence légèrement moins importante est observée lors du passage du modèle de petite taille au modèle de taille moyenne. Cependant, là aussi, une conséquence possible est un manque notable d'entraînement des modèles.

Concentrons-nous sur la figure \ref{fig:whisper2}. Un écart significatif est observé entre deux groupes : les modèles de base de petite taille et les modèles de taille moyenne et grande taille. On constate qu'au sein de ces deux groupes, les modèles donnent approximativement les mêmes résultats pour chaque type d'apprentissage, tandis que la différence entre les métriques finales après 6 époques est assez importante pour le modèle de petite taille. Parallèlement, la métrique entre le premier groupe de modèles et le modèle de petite taille diffère à peu près autant que celle entre le second groupe et le modèle de petite taille. Cela pourrait indiquer une différence qualitative lors du passage aux modèles de taille moyenne et grande taille. En conclusion, dans le modèle E2E ultérieur, il est nécessaire de se concentrer spécifiquement sur le modèle de taille moyenne, car il est nettement supérieur aux autres modèles avec moins de pondérations. Parallèlement, sa qualité n'est pas très inférieure à celle du modèle de grande taille, mais son nombre de pondérations est bien plus élevé (voir le tableau \ref{tb:5}).

De plus, les graphiques plaident en faveur de la quantification des modèles. En effet, l'apprentissage sans quantification est assez lent, de seulement quelques pourcents de la métrique WER. En même temps, même pour les petits modèles, cet apprentissage nécessite beaucoup de temps. Bien sûr, le graphique montre une tendance à la baisse de la métrique, ce qui indique que ces modèles ne sont pas entraînés avec le jeu de données actuel. Cependant, comme mentionné précédemment, leur entraînement complet est difficile. De plus, comme le montrent les graphiques ci-dessus, les modèles affinés utilisant les approches PEFT affichent de meilleurs résultats que les modèles entièrement affinés avec un nombre réduit de paramètres.

Le tableau \ref{tb:7} ci-dessous présente les meilleures valeurs de métrique après 6 périodes d'app\-rentissage pour chaque modèle et chaque type d'apprentissage.

\begin{table}[ht]
 \resizebox{\textwidth}{!}{
 \centering   
\begin{tabular}{|| >{\centering\arraybackslash}m{3,6cm}| >{\centering\arraybackslash}p{2cm} | >{\centering\arraybackslash}p{1cm} | >{\centering\arraybackslash}p{1cm} |>{\centering\arraybackslash}p{1.7cm} |>{\centering\arraybackslash}p{1.7cm}||}
\hline
& réglage fin complet & int4 & int8 & int4 AdaLoRA & int8 AdaLoRA\\
\hline
Minuscule (39 M~paramètres) & \textbf{22.9} & 39.3 & 36.9 & 36.1 & 32.1 \\
\hline
Base (74 M~paramètres) & \textbf{19.0} & 37.4 & 33.5 & 39.9 & 36.5 \\
\hline
Petit (244 M~paramètres) & --- & 31.5 & \textbf{25.4} &30,5 & \textbf{25,1} \\
\hline
Moyen (769 M~paramètres) & --- & 21,1 &15,4 &15,2 & \textbf{11,3} \\
\hline
Grand (1 550 M~paramètres) & --- & 18,8 &11,3 &13,1 & \textbf{9,8} \\
\hline
\end{tabular}
}
\caption{Rendement d'erreur d'exécution (WER) pour différents modèles Whisper affinés après 6 époques d'apprentissage}
\label{tb:7}
\end{table}

Dans le tableau \ref{tb:7}, les meilleurs résultats au sein d'un même modèle de base sont mis en évidence sur chaque ligne. On constate que l'option int8 + AdaLoRA est la plus adaptée aux grands modèles.\section{Conclusion et travaux futurs}

\subsection{Conclusion}

Mon travail présente une nouvelle approche efficace pour la création d'un modèle E2E de résumé audio abstrait. Ce modèle utilise diverses méthodes modernes d'apprentissage automatique : de grands modèles linguistiques et des méthodes de réglage fin des réseaux de neurones.

J'ai étudié les méthodes d'apprentissage LoRA et AdaLoRA pour la tâche de résumé de texte. Quel que soit le modèle de base, la méthode AdaLoRA présente des valeurs supérieures à LoRA et proches de celles d'un modèle entièrement entraîné. L'efficacité de ces méthodes pour la tâche sélectionnée a ainsi été démontrée. De plus, j'ai non seulement étudié le paramètre de compression des adaptateurs par rapport au nombre initial de pondérations, mais j'ai également identifié un hyperparamètre optimal $r$ de la dimension interne de la matrice d'adaptateurs pour chacune des méthodes.

Par ailleurs, au cours de mes travaux, une famille de modèles ASR a été entraînée, qui servira ultérieurement à construire le modèle final de résumé audio E2E. Cette formation a consisté non seulement à obtenir des modèles de qualité comparables aux modèles de pointe, mais aussi à étudier les effets de la quantification et d'AdaLoRA sur le processus d'apprentissage. j'ai comparé les coefficients de compression des modèles, leur temps d'apprentissage et la qualité finale obtenue. De plus, des hyperparamètres ont été sélectionnés pour un apprentissage réussi de ces modèles. L'effet de la qualité d'apprentissage du modèle a également été étudié en fonction de la taille initiale des pondérations. Une différence qualitative a été mise en évidence entre l'apprentissage de petits et de grands modèles, exprimée différemment selon le type d'apprentissage.

En conclusion, un concept a été proposé pour créer un modèle de sommation audio E2E à partir des modèles ci-dessus. Ce modèle devrait présenter de nombreux avantages par rapport à la version existante. D'une part, un tel modèle devrait avoir un poids total plus faible, d'autre part, son temps d'apprentissage devrait être plus court et le processus lui-même devrait nécessiter moins de ressources de calcul.

\subsection{Travaux futurs}

Pendant le reste du stage, l'objectif principal est d'enseigner le modèle E2E final, proposé dans ce travail. De plus, un objectif important est de vérifier les résultats décrits dans l'article, dont je me suis inspirés pour mes travaux.

Je me concentrerai ensuite sur quatre axes d'amélioration de notre modèle. D'une part, je souhaite améliorer la compression du modèle en étudiant l'influence des adaptateurs LoRA, comme cela a été fait dans ce travail \cite{35}. De plus, une amélioration de la qualité peut être obtenue en ajoutant de nouvelles fonctionnalités telles que des images vidéo de l'événement (une approche similaire est utilisée dans cet article \cite{36}), ou en utilisant une approche mixte lorsque le modèle S2T utilise simultanément des approches de résumé abstrait et extractif. D'autre part, je souhaite améliorer la compression du modèle. Cela peut être réalisé grâce à de nouvelles approches, comme dans cet article \cite{34}, où d'autres approches sont proposées pour utiliser simultanément la quantification et les adaptateurs LoRA. Un autre axe de recherche pourrait être d'apprendre à ordonner les entités textuelles pertinentes \cite{CaptneUsu11} avant de résumer le texte. L'ordonnancement peut aider à évaluer la pertinence des différentes parties du texte, facilitant ainsi la sélection des informations les plus importantes à résumer.  Enfin, en organisant les documents en groupes selon leur contenu et leurs thèmes \cite{PESSIOT2010180}, le partitionnement peut améliorer la qualité des résumés.

\bibliographystyle{unsrt}
\bibliography{bibfile}

\begin{thebibliography}{10}

\bibitem{19}
Sourab Mangrulkar, Sylvain Gugger, Lysandre Debut, Younes Belkada, Sayak Paul,
  and Benjamin Bossan.
\newblock Peft: State-of-the-art parameter-efficient fine-tuning methods.
\newblock \url{https://github.com/huggingface/peft}, 2022.

\bibitem{13}
Edward~J. Hu, Yelong Shen, Phillip Wallis, Zeyuan Allen-Zhu, Yuanzhi Li, Shean
  Wang, Lu~Wang, and Weizhu Chen.
\newblock Lora: Low-rank adaptation of large language models, 2021.

\bibitem{14}
Qingru Zhang, Minshuo Chen, Alexander Bukharin, Nikos Karampatziakis, Pengcheng
  He, Yu~Cheng, Weizhu Chen, and Tuo Zhao.
\newblock Adalora: Adaptive budget allocation for parameter-efficient
  fine-tuning, 2023.

\bibitem{15}
Shih-Yang Liu, Chien-Yi Wang, Hongxu Yin, Pavlo Molchanov, Yu-Chiang~Frank
  Wang, Kwang-Ting Cheng, and Min-Hung Chen.
\newblock Dora: Weight-decomposed low-rank adaptation, 2024.

\bibitem{17}
Amir Gholami, Sehoon Kim, Zhen Dong, Zhewei Yao, Michael~W. Mahoney, and Kurt
  Keutzer.
\newblock A survey of quantization methods for efficient neural network
  inference, 2021.

\bibitem{18}
Xunyu Zhu, Jian Li, Yong Liu, Can Ma, and Weiping Wang.
\newblock A survey on model compression for large language models, 2023.

\bibitem{20}
V.~Klema and A.~Laub.
\newblock The singular value decomposition: Its computation and some
  applications.
\newblock {\em IEEE Transactions on Automatic Control}, 25(2):164--176, 1980.

\bibitem{21}
Hugo Touvron, Thibaut Lavril, Gautier Izacard, Xavier Martinet, Marie-Anne
  Lachaux, Timothée Lacroix, Baptiste Rozière, Naman Goyal, Eric Hambro,
  Faisal Azhar, Aurelien Rodriguez, Armand Joulin, Edouard Grave, and Guillaume
  Lample.
\newblock Llama: Open and efficient foundation language models, 2023.

\bibitem{22}
Shayne Longpre, Le~Hou, Tu~Vu, Albert Webson, Hyung~Won Chung, Yi~Tay, Denny
  Zhou, Quoc~V Le, Barret Zoph, Jason Wei, et~al.
\newblock The flan collection: Designing data and methods for effective
  instruction tuning.
\newblock {\em arXiv preprint arXiv:2301.13688}, 2023.

\bibitem{16}
Tim Dettmers, Artidoro Pagnoni, Ari Holtzman, and Luke Zettlemoyer.
\newblock Qlora: Efficient finetuning of quantized llms, 2023.

\bibitem{25}
Bo~He, Jun Wang, Jielin Qiu, Trung Bui, Abhinav Shrivastava, and Zhaowen Wang.
\newblock Align and attend: Multimodal summarization with dual contrastive
  losses, 2023.

\bibitem{23}
Ruipeng Jia, Yanan Cao, Hengzhu Tang, Fang Fang, Cong Cao, and Shi Wang.
\newblock Neural extractive summarization with hierarchical attentive
  heterogeneous graph network.
\newblock In {\em Proceedings of the 2020 Conference on Empirical Methods in
  Natural Language Processing (EMNLP)}, pages 3622--3631, 2020.

\bibitem{24}
Ming Zhong, Pengfei Liu, Yiran Chen, Danqing Wang, Xipeng Qiu, and Xuanjing
  Huang.
\newblock Extractive summarization as text matching, 2020.

\bibitem{amini:hal-01335857}
Massih-Reza Amini and Nicolas Usunier.
\newblock {A Contextual Query Expansion Approach by Term Clustering for Robust
  Text Summarization}.
\newblock In {\em {Document Understanding Conference (DUC)}}, pages 48--55,
  2007.

\bibitem{26}
Massih-Reza Amini.
\newblock {Interactive Learning for Text Summarization}.
\newblock In {\em {PKDD/MLTIA Workshop on Machine Learning and Textual
  Information}}, Lyon, France, 2000.

\bibitem{27}
Matthew Richardson and Pedro Domingos.
\newblock The intelligent surfer: Probabilistic combination of link and content
  information in pagerank.
\newblock 2004.

\bibitem{7}
Roshan Sharma, Shruti Palaskar, Alan Black, and Florian Metze.
\newblock Speech summarization using restricted self-attention.
\newblock 10 2021.

\bibitem{8}
Potsawee Manakul, Mark~J.F. Gales, and Linlin Wang.
\newblock {Abstractive Spoken Document Summarization Using Hierarchical Model
  with Multi-Stage Attention Diversity Optimization}.
\newblock In {\em Proc. Interspeech 2020}, pages 4248--4252, 2020.

\bibitem{28}
Jingqing Zhang, Yao Zhao, Mohammad Saleh, and Peter~J. Liu.
\newblock Pegasus: Pre-training with extracted gap-sentences for abstractive
  summarization, 2020.

\bibitem{29}
Jacob Devlin, Ming-Wei Chang, Kenton Lee, and Kristina Toutanova.
\newblock Bert: Pre-training of deep bidirectional transformers for language
  understanding, 2019.

\bibitem{balikas-amini-2016-twise}
Georgios Balikas and Massih-Reza Amini.
\newblock {T}wi{SE} at {S}em{E}val-2016 task 4: {T}witter sentiment
  classification.
\newblock In {\em Proceedings of the 10th International Workshop on Semantic
  Evaluation ({S}em{E}val-2016)}, pages 85--91, 2016.

\bibitem{10.1007/978-3-031-63536-6_20}
Timon Ziegenbein, Shahbaz Syed, Martin Potthast, and Henning Wachsmuth.
\newblock Objective argument summarization in search.
\newblock In {\em Robust Argumentation Machines: First International
  Conference}, page 335–351, 2024.

\bibitem{Alaa20}
Alaa Mohamed, Mayar Yasser, Mohamed Ayman, Menna Gamil, and Walaa Elashmawi.
\newblock News aggregator and efficient summarization system.
\newblock 11:636--641, 07 2020.

\bibitem{30}
Jesse Dodge, Maarten Sap, Ana Marasović, William Agnew, Gabriel Ilharco, Dirk
  Groeneveld, Margaret Mitchell, and Matt Gardner.
\newblock Documenting large webtext corpora: A case study on the colossal clean
  crawled corpus, 2021.

\bibitem{31}
Kavita Ganesan.
\newblock Rouge 2.0: Updated and improved measures for evaluation of
  summarization tasks, 2018.

\bibitem{32}
Ilya Gusev.
\newblock Dataset for automatic summarization of russian news.
\newblock In {\em Artificial Intelligence and Natural Language}, pages
  {122--134}, Cham, 2020. Springer International Publishing.

\bibitem{33}
Mahnaz Koupaee and William~Yang Wang.
\newblock Wikihow: A large scale text summarization dataset, 2018.

\bibitem{9}
Alec Radford, Jong~Wook Kim, Tao Xu, Greg Brockman, Christine McLeavey, and
  Ilya Sutskever.
\newblock Robust speech recognition via large-scale weak supervision, 2022.

\bibitem{37}
Ramon Sanabria, Ozan Caglayan, Shruti Palaskar, Desmond Elliott, Loïc
  Barrault, Lucia Specia, and Florian Metze.
\newblock How2: A large-scale dataset for multimodal language understanding,
  2018.

\bibitem{35}
Wei Liu, Ying Qin, Zhiyuan Peng, and Tan Lee.
\newblock Sparsely shared lora on whisper for child speech recognition, 2024.

\bibitem{36}
Bin Zhao, Maoguo Gong, and Xuelong Li.
\newblock Audiovisual video summarization, 2021.

\bibitem{34}
Yixiao Li, Yifan Yu, Chen Liang, Pengcheng He, Nikos Karampatziakis, Weizhu
  Chen, and Tuo Zhao.
\newblock Loftq: Lora-fine-tuning-aware quantization for large language models,
  2023.

\bibitem{CaptneUsu11}
Nicolas Usunier, Massih-Reza Amini, and Cyril Goutte.
\newblock Multiview semi-supervised learning for ranking multilingual
  documents.
\newblock In {\em Proceedings of the 2011 European Conference on Machine
  Learning and Knowledge Discovery in Databases}, page 443–458, 2011.

\bibitem{PESSIOT2010180}
Jean-François Pessiot, Young-Min Kim, Massih~R. Amini, and Patrick Gallinari.
\newblock Improving document clustering in a learned concept space.
\newblock {\em Information Processing \& Management}, 46(2):180--192, 2010.

\end{thebibliography}


\end{document}